\documentclass[prb,letterpaper,twocolumn,aps,epsf]{revtex4}
\usepackage{graphicx,amsfonts,epsfig,latexsym,amscd,amsmath,theorem,mathrsfs,color}
\usepackage{epsfig}
\def\comment#1{}

\newcommand{\beg}{\begin{eqnarray}}
\newcommand{\eee}{\end{eqnarray}}

\def\cm#1{}

\newcommand{\I}{{\mathbb{I}}}
\newcommand{\R}{{\mathbb{R}}}
\newcommand{\avec}{\mbox{\boldmath{$A$}}}
\newcommand{\hh}{{\cal H}}
\newcommand{\eps}{\epsilon}
\newcommand{\wt}{\widetilde}
\newcommand{\news}{\setcounter{equation}{0}}

\newcommand{\f}{\frac}
\newcommand{\Ll}{{\cal L}}

\newcommand{\be}{\begin{equation}}
\newcommand{\ee}{\end{equation}}
\newcommand{\ba}{\begin{eqnarray}}
\newcommand{\ea}{\end{eqnarray}}
\newcommand{\beq}{\begin{equation}}
\newcommand{\eeq}{\end{equation}}
\newcommand{\bea}{\begin{eqnarray}}
\newcommand{\eea}{\end{eqnarray}}
\newcommand{\bastar}{\begin{eqnarray*}}
\newcommand{\eastar}{\end{eqnarray*}}

\newcommand{\ra}{\rightarrow}
\newcommand{\cd}{\partial}

\newcommand{\ignore}[1]{}

\begin{document}
\title{Type-1.5 superconductivity in multiband systems: the effects of interband couplings}

\author{
 Johan Carlstr\"om${}^1$, Egor Babaev${}^{1,2}$ and  Martin Speight${}^3$}
\address{
${}^1$Department of Theoretical Physics, The Royal Institute of Technology, Stockholm, SE-10691 Sweden\\
${}^2$ Department of Physics, University of Massachusetts Amherst, MA 01003 USA\\
${}^3$ School of Mathematics, University of Leeds, Leeds LS2 9JT, UK
}

\begin{abstract}
In contrast to single-component superconductors, which are described at
the level of Ginzburg-Landau theory by a single parameter  $\kappa$
and are divided in type-I $\kappa<1/\sqrt{2}$ and type-II
$\kappa>1/\sqrt{2}$ classes, two-component systems in general
possess three fundamental length scales and have been shown to possess a
separate ``type-1.5" superconducting state \cite{bs1,bcs}. In that state, as a
consequence of the extra fundamental length scale, vortices attract one another
at long range but repel at shorter ranges,
and therefore should form
clusters in low magnetic fields. 
In this work we
investigate the appearance of type-1.5
superconductivity and {the interpretation of the} fundamental length scales in the 
case of two active
bands with substantial interband couplings such as intrinsic Josephson
coupling, mixed gradient coupling and density-density interactions. We
show that in the presence of substantial intercomponent interactions
of the above types the system supports type-1.5 superconductivity with
fundamental length scales being associated with the mass of the gauge
field and two masses of normal modes represented  by {\it mixed} combinations of
the density fields.

\end{abstract}
\maketitle
\section{Introduction}
According to Ginzburg-Landau theory, a conventional superconductor near $T_c$
is described by a single complex order parameter field.
 The physics of these systems is governed by two fundamental length scales, 
the magnetic field penetration depth $\lambda$ and the coherence length $\xi$, and the ratio  $\kappa$ of these determines the response to an external field, sorting  them into two categories as follows; type-I when $\kappa <1/\sqrt{2}$ and type-II when $\kappa >1/\sqrt{2}$ \cite{abrikosov}. 

Type-I superconductors expel weak magnetic fields, while strong fields give rise to formation of
 macroscopic normal domains with magnetic flux \cite{prozorov}. The response of type-II superconductors is completely different; below some critical value $H_{c1}$, the field is expelled. Above this value a superconductor forms a lattice 
or a liquid of vortices that have a supercurrent circulating around a normal core and carry magnetic flux through the system. Finally, at a higher second critical value, $H_{c2}$ superconductivity is destroyed. 

These different responses are usually viewed as  consequences of  the vortex interaction in these systems, the energy cost of a boundary between superconducting and normal states and  the thermodynamic stability
of vortex excitations.
In a type-II superconductor the 
energy cost of a boundary between the normal and the superconducting state is 
negative, while  the interaction between vortices is repulsive \cite{abrikosov}.
This leads to a formation of stable vortex lattices and liquids.
In type-I superconductors the situation is the opposite; the vortex interaction is attractive (thus making them
unstable against collapse into one large vortex),
while the boundary energy between normal and superconducting
states is positive. 
From a thermodynamic point of view 
the principal difference between type-I and type-II states is the following:
(i) In type-II superconductors
the external magnetic field strength required to make formation of
 vortex excitations energetically preferred, $H_{c1}$,
is smaller than the thermodynamical
magnetic field $H_{ct}$ (the field 
whose energy density is equal to the condensation energy of a superconductor,
 i.e.\ the field at which the uniform superconducting state
becomes thermodynamically unstable);
(ii) In type-I superconductors the field strength required to create 
a vortex excitation  is larger than the thermodynamical critical magnetic field i.e.\ vortices
cannot form.
One can distinguish also a special ``zero measure" boundary case where $\kappa$
has a critical value exactly at the type-I/type-II boundary,
which in the most common GL model parameterization corresponds to 
 $\kappa = 1/\sqrt{2}$. In that 
case vortices do not interact \cite{remark} in the Ginzburg-Landau theory.

The above circumstances 
result in a situation where, in a strong external magnetic field,
 type-I superconductors
usually have a tendency
to minimize boundary energy between
the normal and superconducting states, leading to a formation of
large inclusions of normal phase which frequently have
laminar structure \cite{prozorov}.

Recently there has been
 increased interest in superconductors with several superconducting components.
The main situations where multiple superconducting components arise are
 (i) multiband superconductors \cite{suhl}-\cite{iron}, (ii)
mixtures of independently conserved condensates such as the projected
superconductivity in metallic hydrogen and hydrogen rich alloys \cite{ashcroft,Nature,herland}
and (iii) superconductors with other than s-wave pairing symmetries.
In this work we focus on the cases (i) and (ii). 
The principal difference between the  cases (i) and (ii)
is the absence of the intercomponent Josephson coupling
in case (ii). 

In two-band superconductors (i) the superconducting components
originate from electronic Cooper pairing in different bands  \cite{suhl}.
Therefore these condensates could not {\it a priori} be expected to be 
 independently conserved.
This, at the level of effective models should manifest itself in a rather
generic presence of intercomponent Josephson coupling.

In the case (ii) two superconducting components were predicted to originate from electronic
and protonic Cooper pairing in metallic hydrogen or hydrogen-rich alloys.
In the projected 
liquid metallic deuterium or deuterium-rich alloys, electronic
superconductivity was predicted to coexist at ultra high pressures with 
deuteronic condensation \cite{Nature,ashcroft,herland}. Because electrons cannot be
converted to protons or deuterons 
the condensates  are independently conserved, 
and therefore in the effective model intercomponent 
Josephson coupling is forbidden on symmetry grounds.
These states are currently a subject of  a renewed experimental pursuit.
They  are expected to arise at high 
but experimentally accessible pressures ($\approx 400  GPa$).
Current static compression experiments achieve pressures of $\approx 350 GPa$ with pressures of an order of $1 TPa$ being anticipated in diamond anvil cell experiments due to the recent availability of ultra hard diamonds. 
Similar two-charged component models were discussed  in the
context of the physics of neutron stars where
they represent  coexistent protonic and $\Sigma^-$-hyperon Cooper pairs in the  {neutron star interior  \cite{ns}.}

This wide variety of systems
raises the need to understand and classify the possible 
magnetic responses of multicomponent superconductors.
It was discussed recently that in multicomponent systems
the magnetic response is much more complex than in ordinary systems,
and that the type-I/type-II dichotomy is not sufficient for classification.
Rather, in a wide range of parameters, as a consequence of the existence
of three fundamental length scales, there is a separate superconducting regime
where vortices have long-range attractive, short-range repulsive interaction
and form vortex clusters immersed in domains of  two-component Meissner state \cite{bs1,bcs}.
Recent experimental works \cite{moshchalkov,moshchalkov2}
have put forward the suggestion that this 
state is realized in
 the two-band material MgB$_2$, which sparked
growing interest in this topic. {In particular
questions were raised over whether this
``type-1.5'' superconducting regime (as it was termed by Moshchalkov 
et al \cite{moshchalkov}, for recent works see \cite{recent})
is possible even in principle in the case of various 
non vanishing couplings (e.g. intrinsic Josephson coupling, mixed gradient couplings etc)
between superconducting components in different bands.}

In this work we  report a study of the appearance of type-1.5
superconductivity especially focusing on the case
of multiband superconductivity, demonstrating the persistence
of this type of superconductivity in the presence of
various kinds of intercomponent couplings (such as interband Josephson coupling, mixed gradient coupling,  density-density,
and other kinds
coupling).

 \subsection{Type-1.5 superconductivity}

The possibility of a new type of superconductivity, distinct from the type-I and type-II 
in multicomponent systems \cite{bs1,bcs}
is based on the following considerations.
In principle the boundary problem  in the Ginzburg-Landau type of equations in the presence 
of phase winding  is  {\it not},
from a rigorous point of view,
 reducible 
to  a one-dimensional problem in general.  
Furthermore, as discussed in \cite{bs1,bcs}, in general
in two-component models  there are {\it three fundamental length scales}
which renders the model impossible to parametrize in terms of a single
dimensionless parameter $\kappa$. 
In the case where the condensates 
are not coupled by interband Josephson coupling but only
by the vector potential 
these length scales are the two independent coherence lengths (set by the
inverse masses of the corresponding scalar density fields)
and magnetic field penetration length (set by the
inverse mass acquired by the gauge field). 
In contrast, in the case where the condensates 
are coupled by interband Josephson terms, one cannot distinguish 
independent coherence lengths attributed to different
condensates. Nonetheless, in this case  the density variations {can also possess two
fundamental length scales \cite{bcs}, in contrast to single-component theories.
We elaborate on this fact below.}
In \cite{bs1,bcs} vortex solutions in two-component theories were found 
which have non-monotonic vortex interaction, with a
long range attractive part
determined by a dominant density-density interaction
and a short range repulsive part produced by current-current and
electromagnetic interactions. 
An important circumstance
which was demonstrated was that these vortices
are thermodynamically stable in spite of the 
existence of the attractive tail in the interaction.

\begin{figure}
\begin{center}
\includegraphics[width=90mm]{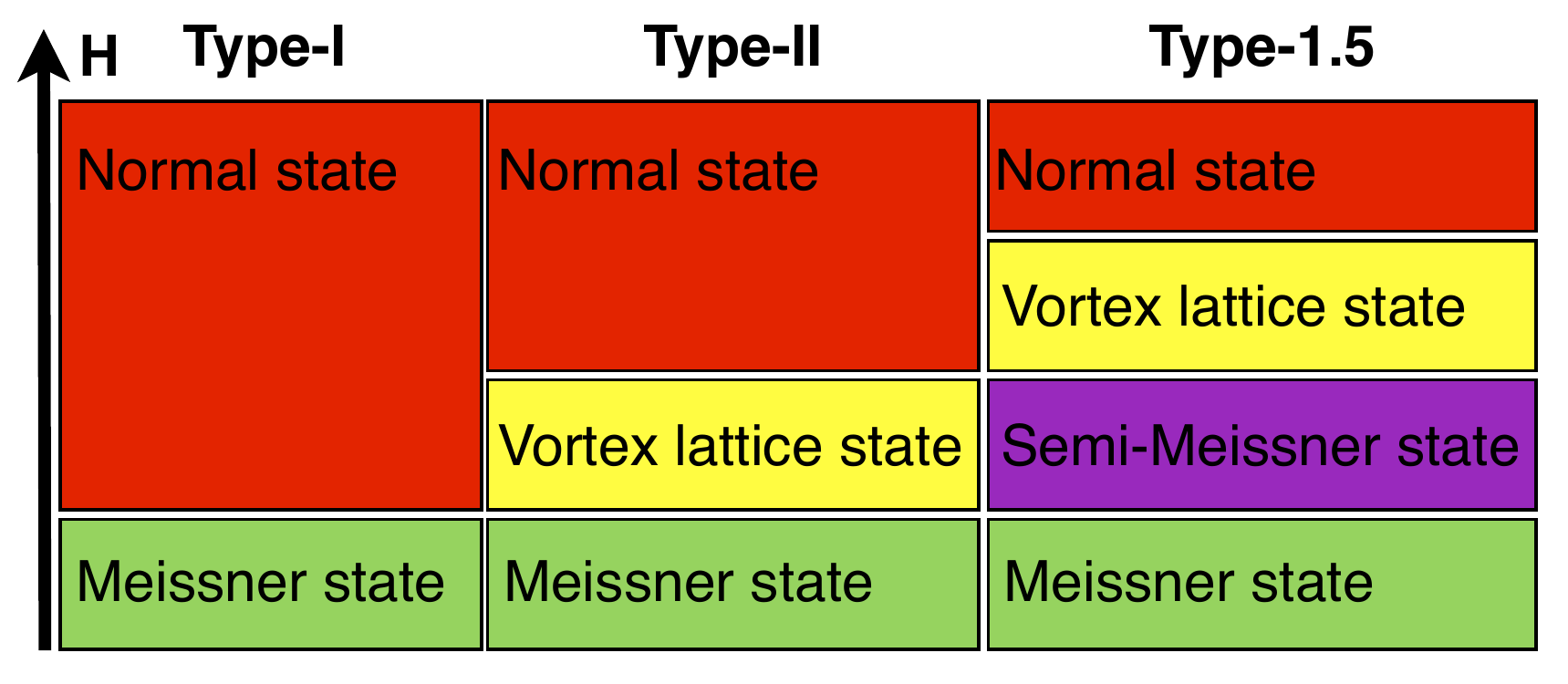}
\end{center}
\caption{ {A comparison of the   magnetic phase diagrams
of clean bulk type-I,type-II and type-1.5 superconductors at zero temperature. The
semi-Meissner state is a macroscopic phase separation into two-component Meissner state
and vortex clusters where one of the density modes is suppressed by core overlaps.}
}
\label{phasediag}
\end{figure}

A non-monotonic intervortex interaction potential should result
in the formation of vortex clusters in low magnetic field
immersed into the vortexless areas, a state
referred to in \cite{bs1} as the ``semi-Meissner state". 
Figure \ref{phasediag} shows the schematic phase diagram
of a type-1.5 superconductor.

If the vortices form clusters one cannot use the usual one-dimensional
argument concerning the energy of superconductor-to-normal state boundary
to classify the magnetic response of the system.
First of all, the energy per vortex in such a case
depends on whether a vortex is placed in a cluster or not:
i.e.\ formation of a single isolated vortex might be energetically
unfavorable, while formation of vortex clusters 
is favorable, because in a cluster where vortices are placed
in a minimum of the interaction potential, the energy 
per flux quantum is smaller than that for an isolated vortex
{ (thermodynamically the nonmonotonic two-vortex interaction potential
predicts that the  smallest energy per flux quantum will
be in the case of a uniform lattice with spacing equal to the minimum
of two-body intervortex potential)}. 

Thus, besides the energy
of a vortex in a cluster, there appears
an additional energy characteristic associated
with the boundary of a cluster.  
In other words, in this situation, to 
determine the magnetic response
of a system 
it is not sufficient to study the one-dimensional boundary problem
nor the single-vortex problem, in contrast to single component systems.
{ Moreover, in a cluster the system tends to minimize the boundary
energy of a cluster (similarly to type-I behavior), while breaking into a lattice of one-quantum vortices
inside the cluster (similarly to type-II systems with negative interface energy).
Thus, in an increased magnetic field the vortices  form via a first order phase transition.
A magnetic phase distinct from the vortex and Meissner states
which then arises is
a macroscopic phase separation into domains of two-component Meissner 
state and vortex clusters where one of the density modes is suppressed 
by core overlap.
We summarize the basic properties of type-I, type-II and type-1.5 regimes in 
the table \ref{table1}.}

The existence of thermodynamically stable type-1.5 superconducting regimes ultimately
depends on the existence of a nonmonotonic intervortex interaction potential.
It is an important question how generic this effect is.
In this work we mainly focus on multiband realizations 
of multicomponent superconductivity and  investigate
 the effects of interband Josephson coupling, mixed gradient coupling, and density-density coupling terms on 
vortex interactions in two band superconductors. We show that (i)
when these couplings are present, the system still
can possess three fundamental length scales, in contrast to the
two length scales in the usual single-component GL theory; (ii) non-monotonic interaction is possible in 
a wide parameter range in these models. 

\begin{table*}
\begin{center}
\begin{tabular}{|p{3cm}||p{4.5cm}|p{4.5cm}|p{5cm}|}
\hline
 & {\bf  single-component Type-I } & {\bf single-component Type-II}  & {\bf multi-component Type-1.5}  \\ \hline \hline
{\bf Characteristic lengths scales} & Penetration length $\lambda$ \&   coherence length $\xi$ ($\frac{\lambda}{\xi}< \frac{1}{\sqrt{2}}$) & Penetration length $\lambda$ \& coherence length $\xi$ ($\frac{\lambda}{\xi}> \frac{1}{\sqrt{2}}$) & Two characteristic density variations length scales $\xi_1$,$\xi_2$ and penetration length $\lambda$,  the nonmonotonic vortex interaction occurs in these systems typically when  $\xi_1<\sqrt{2}\lambda<\xi_2$
\\ \hline 
 {\bf  Intervortex interaction} &	Attractive & 	Repulsive	&Attractive at long range and repulsive at short range \\ \hline
{\bf  Energy of superconducting/normal state boundary} &	Positive	& Negative	 & Under quite general conditions negative energy of superconductor/normal interface inside a vortex cluster but positive energy  of the vortex cluster's boundary  \\ \hline
{\bf The magnetic field required to form a vortex} &	Larger than the thermodynamical critical magnetic field	 & Smaller than thermodynamical critical magnetic field	& In different cases either (i) smaller than the thermodynamical critical magnetic field or (ii) larger than critical magnetic field for single vortex but smaller than critical magnetic field for a vortex cluster of a certain critical size
\\ \hline
{\bf  Phases in external magnetic field } & (i) Meissner state at low fields; (ii) Macroscopically large normal domains at larger fields.
First order phase transition between superconducting (Meissner) and normal states &	(i) Meissner state at low fields, (ii) vortex lattices/liquids at larger fields.  Second order phase transitions between Meissner and vortex states and between vortex and normal states. &	(i) Meissner state at low fields (ii) ``Semi-Meissner state":  vortex clusters coexisting with Meissner domains at intermediate fields (iii) Vortex lattices/liquids at larger fields.  Vortices form via a first order phase transition. The transition from vortex states to normal state is second order.
\\ \hline
{\bf  Energy E(N) of N-quantum axially symmetric vortex solutions} &	$\f{E(N)}{N}$  $<$ $\f{E(N-1)}{N-1}$ for all N. Vortices
coalesce onto a single N-quantum megavortex & 	$\f{E(N)}{N} >\f{E(N-1)}{N-1}$ for all N. N-quantum vortex decays into N infinitely separated single-quantum vortices &	There is a characteristic number N${}_c$ such that $\f{E(N)}{N}$  $<$ $\f{E(N-1)}{N-1}$ for N $<$ N${}_c$, while $\f{E(N)}{N}$ $>$ $\f{E(N-1)}{N-1}$ for N $>$ N${}_c$. N-quantum vortices decay into vortex clusters.
\\ \hline
\end{tabular}
\caption[Basic characteristics of superconductors]{Basic characteristics of bulk clean superconductors in type-I, type-II and type-1.5 regimes. Here the most common units are used in which the value of the GL parameter
which separates type-I and type-II regimes  in a single-component theory is $\kappa_c=1/\sqrt{2}$.}
\label{table1}
\end{center}
\end{table*}


{

The structure of this paper is as follows:
In section II we introduce the model.

In section III  we present a linear
theory of asymptotics of the vortex fields
in a superconductor with two  bands with various
interband couplings.

We begin section  III  by demonstrating that for a {\it  general} 
form of the effective potential in a two-band  (or more generally two-gap) Ginzburg-Landau 
free energy, the linear theory gives, under quite general conditions, two fundamental
length scales of the variations of the densities.
From the linearized theory we calculate the long-range intervortex interaction 
potentials using the two-component generalization of the point-vortex method \cite{spe}
and show how the non-monotonic intervortex interaction potential arise from the
interplay  of two fundamental length scales of the superfluid density variations
and the magnetic field penetration length.
The central point of this part  is how the fundamental length scales are defined in the presence
of interband coupling as well as the occurrence of ``mode mixing".
Next we move to quantitatively study the effects of several   kinds of intercomponent 
couplings which quite generically arise in two-component theories.

In section III (d) we demonstrate that that mixed gradient coupling 
can lead under certain conditions to an {\it increase} in the disparity of 
the characteristic scales of the density variations.

In Section IV  we present a large scale numerical
study of the full nonlinear problem of the  interaction
between a pair of vortices.}

\section{The Model}
\subsection{Free energy functional}

We study the type-1.5 regime using the following two-component Ginzburg-Landau (TCGL) free energy functional.

\begin{eqnarray}
F=\frac{1}{2}(D\psi_1)(D\psi_1)^*+\frac{1}{2}(D\psi_2)(D\psi_2)^*\\ \nonumber
-\nu Re\Big\{(D\psi_1)(D\psi_2)^*\Big\}+\frac{1}{2}(\nabla\times {\bf A})^2 + F_p
\label{gl}
\end{eqnarray}
Here
$D=\nabla + ie {\bf A}$, and $\psi_a=|\psi_a|e^{i\theta_a}$,
$a=1,2$, represent two superfluid components  which,
 in a two-band superconductor
correspond to two superfluid densities in different bands.
{ The term $F_p$ can contain in our analysis an {\it arbitrary} collection of non-gradient terms}.

{ A particular form of two-component GL model which was microscopically derived in \cite{gurevich1,gurevich2,zhitomirsky} for two-band superconductors
is:}

\begin{eqnarray}
F=\frac{1}{2}(D\psi_1)(D\psi_1)^*+\frac{1}{2}(D\psi_2)(D\psi_2)^*\\ \nonumber
-\nu Re\Big\{(D\psi_1)(D\psi_2)^*\Big\}+\frac{1}{2}(\nabla\times {\bf A})^2\\ \nonumber
{+}\alpha_1|\psi_1|^2+\frac{1}{2}\beta_1|\psi_1|^4{+}\alpha_2|\psi_2|^2+\frac{1}{2}\beta_2|\psi_2|^4\\ \nonumber
-\eta_1|\psi_1||\psi_2|\cos(\theta_1-\theta_2)+\eta_2|\psi_1|^2|\psi_2|^2.
\label{gl02}
\end{eqnarray}

The first two terms represent standard Ginzburg-Landau gradient terms,
the second term represents mixed gradient interactions which were shown to originate
in two-band superconductors from impurity scattering  \cite{gurevich1,gurevich2}.
The next term is the magnetic field energy density and the remaining terms represent an effective potential.
Here we note that $\alpha_1$ and $\alpha_2$ can invert sign at different temperatures.
The regime where  $\alpha_1$ is positive while $\alpha_2$ is negative corresponds to the situation 
where one of the bands has no superconductivity of its own but nonetheless bears some superfluid
density due to interband Josephson tunneling,
 which is represented here by the term
$\eta_1|\psi_1||\psi_2|\cos(\theta_1-\theta_2)$. The type-1.5 
behaviour in this regime was studied  in \cite{bcs}.
In this work
we will mainly focus on the situation where both bands are active, i.e.\ $\alpha_{1,2}<0$.
For generality we also add a higher order density-density coupling term $\eta_2|\psi_1|^2|\psi_2|^2$.
We also consider the case of independently conserved condensates where the third and ninth 
terms in (\ref{gl})
are forbidden on symmetry grounds, that is,
$\nu=\eta_1=0$ (see also remark \cite{remark2}).
The equivalence mapping between our units and the standard
textbook units is given in Appendix \ref{appA}.

A microscopic derivation of the TCGL model  (\ref{gl}) requires the fields $|\psi_a|$
to be small. However it does not in principle require $\alpha_a$ to change sign 
at the same temperature. Moreover, as in the case of single-component 
GL theory, we expect the model  (\ref{gl}) to give in many cases a qualitatively acceptable picture
in lower temperature regimes as well.
In fact, our analysis can in some cases give 
a qualitative picture for the case where
one of the fields does not possess
a GL-type effective potential because
the regime where one of the bands is in a London 
limit (i.e. it does not possess a GL effective potential but has a small 
vortex core modeled by a sharp cutoff)
can be recovered from our analysis as a limiting case.
{As will be clear from the analysis below that regime also supports type-1.5 superconductivity.}

\subsection{Basic properties of the vortex excitations.}
The only vortex solutions of the model (\ref{gl}) which have 
finite energy per unit length are the integer $N$-flux quantum vortices
which have the following phase windings 
along a contour $l$  around the vortex core: $\oint_l \nabla \theta_1= 2\pi N,  \oint_l \nabla \theta_2= 2\pi N$.
Vortices with differing phase windings carry a fractional multiple of the
 magnetic
flux quantum and have energy divergent with the system size. These
solutions were investigated in detail in 
\cite{frac}.

In what follows  we investigate only integer flux vortex solutions which 
are the energetically cheapest objects to produce by means of an
external field in a bulk superconductor.

\section{Vortex asymptotics}
\label{asymp}

The key to understanding the interaction of well separated vortices is to 
analyze the large $r$ asymptotics of the vortex solution. We will analyze this
problem in the context of a general TCGL model whose free energy takes the
form
\bea
F&=&\frac12(D_i\psi_1)^*D_i\psi_1+\frac12(D_i\psi_2)^*D_i\psi_2\nonumber \\
&&+\frac12(\cd_1A_2-
\cd_2A_1)^2+F_p
\eea
where $F_p$ contains all the non-gradient terms (in particular,
but not restricted to, Josephson and density-density interaction terms).
This free energy is consistent with (\ref{gl}) in the case $\nu=0$. 
We will show in section \ref{sec:gradgrad} how to handle mixed gradient terms.
{\it The precise form of $F_p$ is
not crucial for our analysis in this section.}
By gauge invariance, it can depend on the
condensates only via $|\psi_1|$, $|\psi_2|$ and (if the condensates are not
 independently conserved) on $\theta_1-\theta_2$. 
We will assume that $F_p$ takes its
minimum value (which we normalize to be $0$) when $|\psi_1|=u_1>0$,
$|\psi_2|=u_2>0$ and $\theta_1-\theta_2=0$. So, either there
is no phase coupling ($F_p$ is independent of $\theta_1-\theta_2$) and the
choice of $\theta_1-\theta_2=0$ is arbitrary, or the phase coupling is such
as to encourage phase locking. (Note that the case of phase anti-locked
fields can trivially be recovered from our analysis by mapping $\psi_2\mapsto
-\psi_2$). 

The field equations are obtained from $F$ by demanding that the total free
energy $E=\int F dx_1dx_2$ is stationary with respect to all variations of
$\psi_1,\psi_2$ and $A_i$. A routine calculation yields
\bea
D_iD_i\psi_a&=&2\frac{\cd F_p}{\cd\psi_a^*}\\
\cd_i(\cd_iA_j-\cd_jA_i)&=&e\sum_{a=1}^2{\rm Im}\left(\psi_a^*D_j\psi_a\right).
\eea
This triple of coupled nonlinear partial differential equations supports solutions of the form
\bea
\psi_a&=&f_a(r)e^{i\theta}\nonumber\\
(A_1,A_2)&=&\frac{a(r)}{r}(-\sin\theta,\cos\theta)\label{ansatz}
\eea
where $f_1,f_2,a$ are real profile functions.
{Note that in some cases mixed gradient terms favour 
non-axially symmetric solutions.}
 In this section we consider 
only axially symmetric vortices.
 Fields within the above ansatz satisfy the field 
equations if and only if the profile functions $f_1(r),f_2(r),a(r)$
satisfy the coupled ordinary differential equation system
\bea
f_a''+\frac{1}{r}f_a'-\frac{1}{r^2}(1+ea)^2f_a&=&\label{5}
\left.\frac{\cd F_p}{\cd |\psi_a|}\right|_{(u_1,u_2,0)}\\\label{6}
a''-\frac{1}{r}a'-e(1+ea)(f_1^2+f_2^2)&=&0.
\eea
The solution we require, the vortex, has boundary behaviour $f_a(r)\ra u_a$,
$a(r)\ra-1/e$ as $r\ra\infty$. So, for large $r$, the quantities
\beq
\eps_a(r)=f_a(r)-u_a,\qquad
\alpha(r)=a(r)+\frac1e
\eeq
are small and so should, to leading order, satisfy the {\em linearization}
of (\ref{5}),(\ref{6}) about $(u_1,u_2,-1/e)$. That is, at large $r$,
\bea
\eps_a''+\frac{1}{r}\eps_a'&=&\sum_{b=1}^2\hh_{ab}\eps_b\label{gm}\\
\alpha''-\frac1r\alpha'-e^2(u_1^2+u_2^2)\alpha&=&0
\eea
where $\hh$ is the Hessian matrix of $F_p(|\psi_1|,|\psi_2|,0)$ about its 
minimum
\beq
\hh_{ab}=\left.\frac{\cd^2F_p}{\cd|\psi_a|\cd|\psi_b|}\right|_{(u_1,u_2,0)}.
\eeq
So $\alpha$ decouples from $\eps_1,\eps_2$ asymptotically, and we see
immediately that
\beq
\alpha(r)=q_0rK_1(\mu_A r),\qquad
\mu_A=e\sqrt{u_1^2+u_2^2}
\eeq
where $K_n$ denotes the $n$th modified Bessel's function of the second 
kind\cite{abrste},
and $q_0$ is an unknown real constant. 
Hence, at large $r$,
\beq\label{cqr}
\avec\sim \left(-\frac{1}{er}+q_0K_1(\mu_A r)\right)(-\sin\theta,\cos\theta).
\eeq
Since, for all $n$, 
\beq
K_n(s)\sim\sqrt{\frac{\pi}{2s}}e^{-s}\quad\mbox{as $s\ra\infty$},
\eeq
it follows that the magnetic field decays exponentially as a function of $r$,
with length scale
(penetration depth)
\beq
\lambda\equiv \frac{1}{\mu_A}=\frac{1}{e\sqrt{u_1^2+u_2^2}}.
\eeq

By contrast, (\ref{gm}) represents, in general, a coupled pair of ordinary differential equations for
$\eps_1,\eps_2$. Since $(u_1,u_2,0)$ is a {\em minimum} of 
$F_p(|\psi_1|,|\psi_2|,\theta_1-\theta_2)$, the
Hessian matrix $\hh$ is a {\em positive definite}
symmetric $2\times 2$ real matrix. Hence its eigenvalues, $\mu_1^2,\mu_2^2$
say, are real and positive, and its eigenvectors, $v_1,v_2$ say, form
an orthonormal basis for $\R^2$. Expanding $\eps=(\eps_1 , \eps_2)^T$ in the
basis $v_1,v_2$
\beq
\eps(r)=\chi_1(r)v_1+\chi_2(r)v_2,
\eeq
we see that $\chi_1,\chi_2$ satisfy the {\em uncoupled} pair of ordinary differential equations
\beq
\chi_a''+\frac1r\chi_a'=\mu_a^2\chi_a,
\eeq
whence
\beq
\chi_a(r)=q_aK_0(\mu_a r)
\eeq
for some (unknown) constants $q_1,q_2$. Since $v_1,v_2$ are orthonormal, there
is an angle $\Theta$, which we call the {\em mixing angle}, such that
the
eigenvectors of $\hh$ are
\beq
v_1=\left(\begin{array}{c}\cos\Theta\\\sin\Theta\end{array}\right),\qquad
v_2=\left(\begin{array}{c}-\sin\Theta\\\cos\Theta\end{array}\right).
\eeq
Hence, at large $r$ the density fields behave as 
\bea
\psi_1&\sim&[u_1+q_1\cos\Theta K_0(\mu_1r)-q_2\sin\Theta K_0(\mu_2r)]e^{i\theta}\nonumber\\
\psi_2&\sim&[u_2+q_1\sin\Theta K_0(\mu_1r)+q_2\cos\Theta K_0(\mu_2r)]e^{i\theta}.\label{alipar}
\eea
where, once again, $K_0$ is a Bessel function.

From this analysis it follows that:
\begin{enumerate}
\item In general there are
three fundamental length scales in the problem (in contrast to the
two length scales
of one-component Ginzburg-Landau theory)
which manifest themselves in the vortex asymptotics, namely $1/\mu_A$,
$1/\mu_1$ and $1/\mu_2$.
\item These are constructed from the vacuum expectation values $u_a$ of
$|\psi_a|$ (in the case of $1/\mu_A$) and from the eigenvalues of $\hh$,
the Hessian matrix of $F_p$ about the vacuum (i.e. the ground state).
\item $1/\mu_A$ can be interpreted as the London penetration length of the 
magnetic field.
\item However, unless the mixing angle $\Theta$ is a multiple of $\pi/2$,
$1/\mu_1$ and $1/\mu_2$ { cannot} be interpreted as the coherence lengths of
$\psi_1,\psi_2$ in the usual  sense. This is because the normal modes of the field theory
close to the vacuum are not $|\psi_a|-u_a$, but rather
\bea
\chi_1&=&(|\psi_1|-u_1)\cos\Theta-(|\psi_2|-u_2)\sin\Theta\nonumber \\
\chi_2&=&(|\psi_1|-u_1)\sin\Theta-(|\psi_2|-u_2)\cos\Theta\nonumber 
\eea
obtained by rotating through the mixing angle $\Theta$,
which is also determined by $\hh$.
Therefore in general 
(e.g. in the presence of  intercomponent Josephson coupling)
for a one-flux quantum axially symmetric vortex,  the recovery 
of both fields $\psi_a$ at {\it very} long range
will be according to the same exponential law, set by the smaller of
the masses $\mu_1,\mu_2$; {One should use the representation in terms
of the fields $\chi_{1,2}$ to be handle properly the 
two length scales associated with the density recovery.}
\item This analysis tells us only about the vortex structure at large $r$.
It gives no direct information on the vortex core, which is important to
understand quantitatively the nature of the vortex interactions at intermediate and short distances
which will be studied numerically in the section V.
\end{enumerate}

Since the gauge field mediates a repulsive force between vortices, while
the condensate fields mediate an attractive force, it is clear that we can
read off from the above analysis the condition under which the
intervortex force is attractive at long range: we require that $1/\mu_A$
is {\em not} the longest of the three length scales, or, more
explicitly, that (at least) one of the eigenvalues of $\hh$ should be
less that $\mu_A^2=e^2(u_1^2+u_2^2)$.  We can predict an explicit formula for the long
range two-vortex interaction potential, using the point vortex
formalism \cite{spe} ({a brief review of the method in given in Appendix \ref{appC})}. This rests on the 
observation that, far from its core, the fields of the vortex are
identical to those of a hypothetical point particle
in a linear theory with two Klein-Gordon fields ($\chi_1$ and $\chi_2$ above)
of mass $\mu_1,\mu_2$ and a vector field ($A$) of mass $\mu_A$. The 
point particle carries scalar monopole charges $2\pi q_1$ and $2\pi q_2$
and a magnetic dipole moment $2\pi q_0$. Two such hypothetical particles
held distance $r$ apart would experience an interaction potential
\beq
V(r)=2\pi\left[q_0^2K_0(\mu_A r)-q_1^2K_0(\mu_1r)-q_2^2K_0(\mu_2r)\right].
\label{interact}
\eeq
This formula reproduces the prediction explained above: the long range 
interaction will be attractive if (at least) one of $\mu_1,\mu_2$ is less
than $\mu_A$. 

{ One can ask, retrospectively, whether the approximation of {\em linearizing} in the 
small quantities $\alpha(r),\chi_1(r),\chi_2(r)$ is well justified. Rigorous analysis
of the single component model \cite{plohr} shows that if either of the scalar mode masses,
$\mu_2$ say, exceeds $2\mu_A$, then quadratic terms
in $\alpha$ become comparable at large $r$ with linear terms in $\chi_2$, so that
the equation for $\chi_2$ should include extra terms. In this case, $\chi_2$ decays
like $K_0(\mu_Ar)^2$ rather than $K_0(\mu_2 r)$. One should note, however that, unless
$\mu_1>2\mu_A$ also, the {\em leading} term in (\ref{alipar}), decaying like $K_0(\mu_1 r)$, is
still correct, and it is only the leading term which
determines the nature (attractive or repulsive) of the intervortex interactions at long range.
The case of interest to us is when the long-range force is attractive, that is, when at
least one of $\mu_1,\mu_2$ is {\em less than} $\mu_A$, so the linearized analysis presented
above suffices for our purposes.}

\subsection{The  $U(1)\times U(1)$ symmetric model}\label{basic}

We first illustrate the above analysis in the case
of $U(1)\times U(1)$ condensates coupled
only by a gauge field  \cite{bs1}
where
\beq
F_p=\alpha_1|\psi_1|^2+\frac{\beta_1}{2}|\psi_1|^4+
\alpha_2|\psi_2|^2+\frac{\beta_2}{2}|\psi_2|^4+\mbox{constant},
\label{cuqura}\eeq
{with both $\alpha_1<0$ and $\alpha_2<0$}.
Here $F_p$ is independent of $\theta_1-\theta_2$ and its minimum occurs
at $|\psi_a|=u_a$ where
\beq
u_a=\sqrt{\frac{-\alpha_a}{\beta_a}}.
\eeq
The Hessian matrix of $F_p$ at $(u_1,u_2,0)$ is
\beq
\hh=\left(\begin{array}{cc}-4\alpha_1&0\\0&-4\alpha_2\end{array}\right)
\eeq
whose eigenvalues are $\mu_a^2=-4\alpha_a$, with corresponding
eigenvectors $v_1=(1,0)^T$, $v_2=(0,1)^T$. Hence, the mixing angle is 
$\Theta=0$, the penetration depth is
\beq
\lambda \equiv \frac{1}{\mu_A}=e^{-1}\left(\frac{-\alpha_1}{\beta_1}+
\frac{-\alpha_2}{\beta_2}\right)^{-\frac12}
\eeq
and the density decay lengths are the usual coherence lengths
\beq
\xi_a\equiv \frac{1}{\mu_a}=\frac{1}{2\sqrt{-\alpha_a}}.
\eeq
The criterion for long-range vortex attraction therefore
amounts to the requirement that one of the coherence lengths is
larger than the magnetic field penetration length,
\beq
\min\{2\sqrt{-\alpha_1},2\sqrt{-\alpha_2}\}<e\sqrt{\frac{-\alpha_1}{\beta_1}+
\frac{-\alpha_2}{\beta_2}}.
\eeq
Note  this criterion indicates only a long range attraction.
For a realization of the type-1.5 regime one should additionally require
short range repulsion and thermodynamic stability of a vortex cluster 
(i.e.\ a vortex cluster should become energetically favorable to form in 
external fields smaller than the thermodynamical critical magnetic field).

{Note also that, in systems with independently conserved
condensates, such GL models arise from expansion of the free energy
in powers $(1-T/T_{c1})$ and  $(1-T/T_{c 2})$ near two   critical temperatures (such systems are, for example,
 the projected superconducting states of metallic hydrogen
or condensate mixtures in neutron stars). In general $T_{c1}$ and $T_{c2}$ can be quite different
making two such expansions at the same temperature formally  impossible. In that case the  more suitable model is  a London model for one component  (i.e. $|\psi_1| \approx  const$ except a  core cutoff) coupled to a GL model of the second component. When it is the component with ``short" coherence
length   which is modeled by the London limit,
we recover a description of the type-1.5 behavior in that case from our above analysis
as a simple limit.

\subsection{Josephson coupling}
\label{joco}

We next consider how this picture is influenced by the addition
of an interband Josephson term which breaks the  $U(1) \times U(1)$ symmetry to $U(1)$ (note that this
in particular implies that there is only one critical temperature),
\beq
F_p=\hat{F}_p-\eta_1|\psi_1||\psi_2|\cos(\theta_1-\theta_2),
\eeq
where $\hat{F}_p$ is the free energy
defined in (\ref{cuqura}) and  the Josephson coupling $\eta_1>0$, so that
$F_p$ is minimized when $\theta_1-\theta_2=0$. 
Adding this term changes the vacuum expectation
values $u_a$ of the fields.  To find $u_1,u_2$ we must solve
\beq
\frac{\cd F_p}{\cd|\psi_1|}=\frac{\cd F_p}{\cd|\psi_2|}=0,
\eeq
that is,
\bea\label{a1}
2\alpha_1u_1+2\beta_1u_1^2&=&\eta_1 u_2\\\label{a2}
2\alpha_2u_2+2\beta_2u_2^2&=&\eta_1 u_1.
\eea
Unfortunately, it is not possible to solve these equations explicitly, 
except in special cases. For particular values of
the parameters $\alpha_a,\beta_a,\eta_1$ they can easily be solved numerically,
as can the eigenvalue problem for $\hh$. Note that 
like in the case of uncoupled bands, there are in general three 
fundamental length scales also in the presence of the Josephson term 
which can then
be computed. We present numerical analysis of this problem in the full Ginzburg-Landau model in section
{\ref{numerics}}.
To make analytical advance in this section we
 treat the $\eta_1$ dependence of the length scales perturbatively.  That is,
we will construct Taylor expansions for $u_a(\eta_1)$, $\mu_A(\eta_1)$,
$\mu_a(\eta_1)$ and $\Theta(\eta_1)$. To keep the presentation simple, we
will work to order $\eta_1^1$, so the results will give the leading correction to
the formulae of the previous section as the Josephson coupling $\eta_1$ is
``turned on''. Higher order corrections are easily computed but do not
give much extra insight.

Let us denote quantities defined in the uncoupled model (at $\eta_1=0$)
with a hat, so $\hat{u}_a=\sqrt{-\alpha_a/\beta_a}$ are the
uncoupled vacuum expectation values of $|\psi_a|$, for example. Let 
$u(\eta_1)=(u_1(\eta_1),u_2(\eta_1))^T$ and 
\beq
G(|\psi_1|,|\psi_2|)=\left(\begin{array}{c}
\cd F_p/\cd|\psi_1|\\\cd F_p/\cd|\psi_2|\end{array}\right).
\eeq
Then, by definition $G(u(\eta_1))=0$ for all $\eta_1$, and $\hat{G}(\hat{u})=0$.
Differentiating with respect to $\eta_1$ (denoted by a prime), 
we see that 
\bea
0&=&\frac{\cd G}{\cd\eta_1}(\hat{u})+\hat{\hh}{u'}(0)\nonumber\\
\Rightarrow
{u'}(0)&=&-\hat{\hh}^{-1}\frac{\cd G}{\cd\eta_1}(\hat{u})\nonumber\\
&=&-\left(\begin{array}{cc}\hat\mu_1^{-2}&0\\0&\hat\mu_2^{-2}\end{array}\right)
\left(\begin{array}{c}-\hat{u}_2\\-\hat{u}_1\end{array}\right).
\eea
Hence the ground state densities receive a correction linear in $\eta_1$
\bea
u_1(\eta_1)&=&\hat{u}_1+\frac{\hat{u}_2}{\hat{\mu}_1^2}\eta_1+O(\eta_1^2)\nonumber\\
u_2(\eta_1)&=&\hat{u}_2+\frac{\hat{u}_1}{\hat{\mu}_2^2}\eta_1+O(\eta_1^2).
\eea
From this expression for ground state densities one can readily
calculate the gauge field mass $\mu_A(\eta_1)$, whose inverse  gives the London
penetration length. One sees that
\beq
\mu_A(\eta_1)^2=\hat{\mu}_A^2+2\hat{u}_1\hat{u}_2\left(\frac{1}{\hat\mu_1^2}
+\frac{1}{\hat\mu_2^2}\right)\eta_1+O(\eta_1^2).
\eeq
So the effect of the Josephson coupling is always to increase the vacuum
expectation values of $|\psi_a|$, and hence to decrease the penetration
depth $1/\mu_A$.

The other two length scales are the eigenvalues of $\hh$ where
\bea
\hh_{ab}&=&\left.\frac{\cd^2\:\:\:}{\cd|\psi_a|\cd|\psi_b|}(\hat{F}_p
-\eta_1|\psi_1||\psi_2|)\right|_{u(\eta_1)}\nonumber\\
&=&\hat\hh_{ab}+\eta_1\left.
\frac{\cd^3\hat{F}_p}{\cd|\psi_a|\cd|\psi_b|\cd|\psi_c|}\right|_{\hat{u}}
{u'}_c(0)\nonumber\\
&&-\eta_1(1-\delta_{ab})+O(\eta_1^2).
\eea
Now
\beq
\frac{\cd^3\hat{F}_p}{\cd|\psi_a|\cd|\psi_b|\cd|\psi_c|}=\left\{
\begin{array}{ll}12\beta_a|\psi_a|&\mbox{if $a=b=c$}\\
0&\mbox{otherwise}\end{array}\right.
\eeq
so 
\bea
\hh&=&\hat\hh+\eta_1\left(\begin{array}{cc}
12\beta_1\hat{u}_1{u'}_1(0)&-1\\
-1&12\beta_2\hat{u}_2{u'}_2(0)\end{array}\right)+O(\eta_1^2)\nonumber \\
&=&\left(\begin{array}{cc}
\hat{\mu}_1^2+3\eta_1\hat{u}_2/\hat{u}_1&-\eta_1\\-\eta_1&\hat\mu_2^2+3\eta_1\hat{u}_1/\hat{u}_2\end{array}\right)+O(\eta_1^2).
\eea
So the eigenvalues $\lambda=\mu_a^2$ satisfy the characteristic
equation
\beq
\left(\hat\mu_1^2+3\eta_1\frac{\hat{u}_2}{\hat{u}_1}-\lambda\right)
\left(\hat\mu_2^2+3\eta_1\frac{\hat{u}_1}{\hat{u}_2}-\lambda\right)+O(\eta_1^2)=0.
\eeq
Hence
\bea
\mu_1^2&=&\hat\mu_1^2+3\eta_1\frac{\hat{u}_2}{\hat{u}_1}+O(\eta_1^2)\nonumber \\
\mu_2^2&=&\hat\mu_2^2+3\eta_1\frac{\hat{u}_1}{\hat{u}_2}+O(\eta_1^2).
\label{mus}
\eea
As with the penetration depth, the effect of Josephson coupling (at leading 
order), is to {\em decrease} the characteristic length scales $1/\mu_a$
of both normal modes.
Another effect of Josephson coupling is mixing the fields.
Let us now compute the corresponding mixing angle $\Theta(\eta_1)$. Recall that this is,
by definition, the angle such that 
\beq
v(\eta_1)=\left(\begin{array}{c}\cos\Theta(\eta_1)\\\sin\Theta(\eta_1)\end{array}
\right)
\eeq
is the eigenvector of $\hh$ with eigenvalue $\mu_1^2$. We know that
$v(0)=\hat{v}=(1,0)^T$, that $v(\eta_1)\cdot v(\eta_1)=1$ and that
\beq\label{gear}
M(\eta_1)v(\eta_1)=0,\qquad \mbox{where } M(\eta_1)=\hh(\eta_1)-\mu_1(\eta_1)^2\I_2
\eeq
for all $\eta_1$. As computed above,
\beq
M(\eta_1)=\left(\begin{array}{cc}0&0\\0&\hat\mu_2^2-\hat\mu_1^2\end{array}\right)
+\eta_1\left(\begin{array}{cc}0&-1\\-1&3(\frac{\hat{u}_1}{\hat{u}_2}-
\frac{\hat{u}_2}{\hat{u}_1})\end{array}\right)+O(\eta_1^2)
\eeq
Differentiating (\ref{gear}) and $v(\eta_1)\cdot v(\eta_1)=1$
with respect to $\eta_1$ yields
\beq
{M}'(0)\hat{v}+M(0){v'}(0)=0,\qquad v(0)\cdot{v'}(0)=0
\eeq
which can be solved for ${v'}(0)$. One finds that
\beq
v(\eta_1)=\left(\begin{array}{c}1\\0\end{array}\right)+\eta_1
\left(\begin{array}{c}0\\(\hat\mu_2^2-\hat\mu_1^2)^{-1}\end{array}\right)+
O(\eta_1^2).
\eeq
Hence, the mixing angle is
\beq
\Theta(\eta_1)=\frac{\eta_1}{\hat\mu_2^2-\hat\mu_1^2}+O(\eta_1^2).
\eeq
Thus the Josephson term produces mode mixing. Clearly, this perturbative
expansion is well-defined only if $\hat\mu_1\neq\hat\mu_2$. In the
case where $\hat\mu_1=\hat\mu_2$, $\hat\hh=\hat\mu_1^2\I_2$ and the
assertion that $v(0)=(1,0)^T$ is arbitrary: any orthonormal pair of
vectors can be taken as the eigenvectors associated with $\hat\mu_1^2,
\hat\mu_2^2$. Hence, the notion of ``mixing angle" is ill-defined in this
case, and is not amenable to perturbative calculation.

The normal modes are associated with the following combinations of the $|\psi_a|$ 
fields
\bea
\chi_1&=&(|\psi_1|-u_1)\cos\biggl[\frac{\eta_1}{\hat\mu_2^1-\hat\mu_1^2}\biggr]-(|\psi_2|-u_2)\sin\biggl[\frac{\eta_1}{\hat\mu_2^1-\hat\mu_1^2}\biggr]\nonumber \\
\chi_2&=&(|\psi_1|-u_1)\sin\biggl[\frac{\eta_1}{\hat\mu_2^1-\hat\mu_1^2}\biggr]-(|\psi_2|-u_2)\cos\biggl[\frac{\eta_1}{\hat\mu_2^1-\hat\mu_1^2}\biggr].\nonumber 
\eea
Thus one can associate  ``coherence lengths" 
of these fields 
 with the inverse masses of the model (\ref{mus})
 which are functions of the coherence lengths ($\hat\mu_a^{-1}$) 
and vacuum field densities ($\hat{u}_a$) defined in the
 Josephson-uncoupled theory, and the  strength 
 of the Josephson coupling $\eta_1$.
 Note that, returning to the original fields $|\psi_a|$, the
very long-range behavior of both of these density fields
is governed by whichever of the fields $\chi_{1,2}$ has the slower
recovery rate. This  implies that at very long range
both fields $|\psi_a|$ should have the {\it same} exponential recovery law
set by the smaller of $\mu_a$. The physical meaning
of mode mixing is that the
variation of the original density fields $|\psi_a|$
acquires two length scales and one should rotate the {fields} through the
mixing angle $\Theta$ to determine the normal modes $\chi_{1,2}$
whose recovery rates are governed by different single exponential laws.
 
{ Another point  to note here is that,
from a quantitative point of view, turning on a small  Josephson coupling does not
radically alter the integer flux  vortices. For small $\eta_1$, there is 
a small correction to each of the length scales (all three length scales
become smaller), and there is a small amount of normal mode mixing 
(measured by $\Theta(\eta_1)$).} Therefore for composite integer flux vortices the 
addition of
Josephson coupling does not represent any kind of singular perturbation.
\subsubsection{{Comparison with the case of passive second  band}}
It is interesting to compare these results with the case where one of the bands,
$\psi_2$ say, is passive, and has superconductivity only by virtue of
the Josephson coupling term\cite{bcs}. In this case, the free energy $F_p$
has { $\alpha_2>0$}. In the uncoupled model (at $\eta_1=0$), $F_p$ is minimized
when $|\psi_1|=\hat{u}_1=\sqrt{-\alpha_1/\beta_1}$ and $|\psi_2|=\hat{u}_2=0$.
The gauge field has mass $\hat{\mu}_A=e\hat{u}_1$, there is no
mode mixing, and the condensates have masses $\hat{\mu}_1=2\sqrt{-\alpha_1}$
and $\hat{\mu}_2=\sqrt{2\alpha_2}$. The vortex solution has $\psi_2=0$
everywhere and is identical to the Abrikosov vortex of the
single component 
GL theory with $\psi_2$ set to zero. As the Josephson coupling is turned
on, $\psi_2$ acquires a vacuum density of order $\eta_1^1$, mode-mixing
develops, and the three length scales acquire corrections. Repeating the
arguments of the $\alpha_2<0$ case, one finds that
\bea
u_1(\eta_1)&=&\hat{u}_1+O(\eta_1^2)\nonumber\\
u_2(\eta_1)&=&\frac{\hat{u}_1}{\hat{\mu}_2^2}\eta_1+O(\eta_1^2)\nonumber\\
\mu_A(\eta_1)^2&=&\hat{\mu}_A^2+O(\eta_1^2)\nonumber\\
\mu_1(\eta_1)^2&=&\hat{\mu}_1^2+O(\eta_1^2)\nonumber\\
\mu_2(\eta_1)^2&=&\hat{\mu}_2^2+O(\eta_1^2)\nonumber\\
\Theta(\eta_1)&=&\frac{\eta_1}{\hat{\mu}_2^2-\hat{\mu}_1^2}+O(\eta_1^2).
\eea
A significant difference from the case of two active bands {($\alpha_2<0$)}
is that all three of the length scales receive corrections only at order
$\eta_1^2$. Nevertheless, there is mode-mixing at order $\eta_1^1$.

\subsection{Density-density coupling}
\label{densdens}

A similar perturbative analysis of the case when there is 
bi-quadratic density-density coupling
\beq
F_p=\hat{F}_p+\frac{\eta_2}{2}|\psi_1|^2|\psi_2|^2
\eeq
can be carried out. Since the calculations are similar, we merely record the 
results (once again, hatted parameters refer to the uncoupled
$\eta_2=0$ model):
\bea
u_1(\eta_2)&=&\hat{u}_1+\frac{\hat{u}_1\hat{u}_2^2}{\hat\mu_1^2}\eta_2+O(\eta_2^2)
\nonumber\\
u_2(\eta_2)&=&\hat{u}_2+\frac{\hat{u}_2\hat{u}_1^2}{\hat\mu_2^2}\eta_2+O(\eta_2^2)
\nonumber\\
\mu_A(\eta_2)^2&=&\hat\mu_A^2+2\hat{u}_1^2\hat{u}_2^2\left(\frac{1}{\hat\mu_1^2}
+\frac{1}{\hat\mu_2^2}\right)\eta_2+O(\eta_2^2)\nonumber \\
\mu_1(\eta_2)^2&=&\hat\mu_1^2+\left(1+3\frac{\hat{u}_2}{\hat{\mu}_1^2}\right)
\hat{u}_2^2\eta_2+O(\eta_2^2)\nonumber\\
\mu_2(\eta_2)^2&=&\hat\mu_2^2+\left(1+3\frac{\hat{u}_1}{\hat{\mu}_2^2}\right)
\hat{u}_1^2\eta_2+O(\eta_2^2)\nonumber\\
\Theta(\eta_2)&=&\frac{2\hat{u}_1\hat{u}_2}{\hat\mu_1^2-\hat\mu_2^2}\eta_2
+O(\eta_2^2).
\eea
The effect of the extra term is to reduce (if $\eta_2>0$)
or increase (if $\eta_2<0$) all three length scales and to introduce  
 mode mixing.  We present numerical analysis of this kind of coupling in section {\ref{numerics}}.

\subsection{Mixed gradient terms}
\label{sec:gradgrad}

In this section we consider the case where the free energy has 
gradient-gradient coupling terms,
\bea
F&=&\frac12(D_i\psi_1)^*D_i\psi_1+\frac12(D_i\psi_2)^*D_i\psi_2\nonumber\\
&&-\frac\nu2[
(D_i\psi_1)^*D_i\psi_2+(D_i\psi_2)^*D_i\psi_1]+F_p
\eea
where $F_p(|\psi_1|,|\psi_2|,\theta_1-\theta_2)$ is, as before, a non-negative 
function minimized at $(u_1,u_2,0)$.
 In contrast to the previous
two cases, this we can treat exactly, without resorting to power series 
expansion in the coupling parameter $\nu$. We can assume $\nu>0$ without 
loss of generality 
(the case $\nu<0$ is obtained by mapping $\psi_2\mapsto-\psi_2$), and
we must have $\nu<1$, or else $F$ is not positive definite. 

This
case does not fit into the general analysis presented above.
Nonetheless a similar method, with some modification, can be applied.

The field equations are
\bea
D_iD_i(\psi_1-\nu\psi_2)&=&2\frac{\cd F_p}{\cd\psi_1^*}\\
D_iD_i(\psi_2-\nu\psi_1)&=&2\frac{\cd F_p}{\cd\psi_2^*}\\
\cd_i(\cd_iA_j-\cd_jA_i)&=&e\, {\rm Im}\left(\psi_1^*D_j(\psi_1-\nu\psi_2)
\right.\nonumber\\
&&\quad+\left.\psi_2^*D_j(\psi_2-\nu\psi_1)\right)
\eea
which support vortex solutions of the form (\ref{ansatz}) provided the
profile functions obey the coupled system
\bea
&&\left(\frac{d^2\:}{dr^2}+\frac1r\frac{d\:}{dr}-\frac{1}{r^2}(1+ea)^2\right)P
\left(\begin{array}{c}f_1\\f_2\end{array}\right)\nonumber\\
&&\qquad\qquad\qquad\qquad\qquad\qquad=
\left.\left(\begin{array}{c}\cd F_p/\cd|\psi_1|\\\cd F_p/\cd|\psi_2|\end{array}\right)\right|_{(f_1,f_2,0)}\nonumber\\
&&a''-\frac1ra'-e(1+ea)(f_1^2-2\nu f_1f_2+f_2^2)=0,
\eea
where
\beq\label{gimo}
P=\left(\begin{array}{cc}1&-\nu\\-\nu&1\end{array}\right).
\eeq
The vortex boundary conditions are $f_a(r)\ra u_a$
and $a(r)\ra-1/e$ as $r\ra\infty$. Note that these are independent of
$\nu$. We again define $\eps(r)=(f_1(r)-u_1,f_2(r)-u_2)^T$ and $\alpha(r)=a(r)
+1/e$, and linearize the system about $\eps_1=\eps_2=\alpha=0$:
\bea
\left(\frac{d^2\:}{dr^2}+\frac1r\frac{d\:}{dr}\right)P\eps&=&\hh\eps\nonumber\\
\alpha''-\frac1r\alpha'-e^2(u_1^2-2\nu u_1u_2+u_2^2)\alpha&=&0.
\eea
We immediately see that $\avec$ behaves asymptotically as in (\ref{cqr}), but 
with
\beq\label{gilmor}
\mu_A=e\sqrt{u_1^2-2\nu u_1u_2+u_2^2}.
\eeq
The effect of the gradient-gradient coupling is thus to {\em increase}
the penetration depth $1/\mu_A$. Note that the effect would
be opposite if there is a competing sufficiently strong Josephson term which 
enforces phase {\em anti-locking} in the vacuum, that is
phase difference $\theta_1-\theta_2=\pi$.

To decouple the pair of equations for $\eps_1,\eps_2$ we must expand $\eps$ 
in a basis of eigenvectors, not of $\hh$, but rather of
\beq
\wt\hh(\nu)=P(\nu)^{-1}\hh.
\eeq
Note that this matrix is {\em not} in general symmetric. Nonetheless, it can
be shown that its eigenvalues are real and positive for all $0\leq\nu<1$
(see the appendix B). Let $\mu_1^2,\mu_2^2$ be the eigenvalues of $\wt{\hh}$ and
$v_1,v_2$ be the corresponding eigenvectors. Then the condensate fields at large
$r$ take the form
\beq
\left(\begin{array}{c}\psi_1\\\psi_2\end{array}\right)\sim
\left\{\left(\begin{array}{c}u_1\\u_2\end{array}\right)+
q_1K_0(\mu_1r)v_1+q_2K_0(\mu_2r)v_2\right\}e^{i\theta}
\label{pmg}
\eeq
Once again, the condition for long-range attraction is
\beq
\min\{\mu_1^2,\mu_2^2\}<\mu_A^2
\eeq
but now $\mu_1^2,\mu_2^2$ are the eigenvalues of $P(\nu)^{-1}\hh$, not $\hh$,
and $\mu_A$ depends on $u_1,u_2$ and $\nu$, as in (\ref{gilmor}).
Note that
\beq
\mu_1^2\mu_2^2=\det\wt\hh=\det\hh/\det P(\eta_1)=\frac{\det\hh}{1-\nu^2}
\eeq
so the effect of the coupling must be to increase (at least) one of
$\mu_{1,2}$ and hence to decrease (at least) one of the normal mode recovery length
scales. 
Since $\wt\hh$ is not symmetric,
there is no reason why $v_1$, $v_2$ should be orthogonal, so it is not possible
to define a single mixing angle in this case.

To illustrate, consider the simplest case, where $F_p$ is defined as in
(\ref{cuqura}). Then $u_a=\sqrt{-\alpha_a/\beta_a}$ and the uncoupled
$\nu=0$ model has $\hat{\mu}_a^2=-4\alpha_a$ and 
$\hat\mu_A^2=e^2(u_1^2+u_2^2)$. For $\nu>0$, the vacuum expectation 
values do not change, but the penetration depth increases, since
\be\label{heug}
\mu_A^2=\hat\mu_A^2-2\nu u_1u_2.
\ee
 Furthermore
\beq
\wt{\hh}(\nu)=P(\nu)^{-1}\left(\begin{array}{cc}
\hat\mu_1^2&0\\0&\hat\mu_2^2\end{array}\right)=
\frac{1}{1-\nu^2}\left(\begin{array}{cc}\hat\mu_1^2&\nu\hat\mu_2^2\\
\nu\hat\mu_1^2&\hat\mu_2^2\end{array}\right)
\eeq
whose eigenvalues are
\beq
\mu_{1,2}^2(\nu)=\frac{1}{2(1-\nu^2)}\left(
\hat\mu_1^2+\hat\mu_2^2\pm\sqrt{(\hat\mu_1^2-\hat\mu_2^2)^2+4\nu^2
\hat\mu_1^2\hat\mu_2^2}\right).
\eeq
{Now $\mu_{1,2}^{-1}(\nu)$ are the new fundamental length scales
which control the variation of the density fields (\ref{pmg}).}
Without loss of generality, we may assume that $\hat\mu_1\geq\hat\mu_2$ (if 
$\hat\mu_1<\hat\mu_2$ then we simply swap the labels of the condensates). In 
this case, for $\nu>0$ it is clear from the above expression that
\beq\label{ge1ar1}
\mu_1^2>\frac{1}{2(1-\nu^2)}\left(
\hat\mu_1^2+\hat\mu_2^2+\sqrt{(\hat\mu_1^2-\hat\mu_2^2)^2}\right)=
\frac{\hat\mu_1^2}{1-\nu^2}
\eeq
so when $0<\nu<1$, $\mu_1(\nu)>\hat\mu_1$. Recall that 
\beq
\mu_1^2\mu_2^2=\det\wt\hh=\frac{\hat\mu_1^2\hat\mu_2^2}{1-\nu^2},
\eeq
so
\beq
\mu_2^2=\frac{\hat\mu_1^2}{\mu_1^2}\frac{\hat\mu_2^2}{1-\nu^2}
<\hat\mu_2^2
\eeq
by (\ref{ge1ar1}), and hence $\mu_2(\nu)<\hat\mu_2$ when $0<\nu<1$.
{\it In this case, the effect of
gradient-gradient coupling is to decrease the smaller of the 
normal mode decay lengths, $\mu_1^{-1}$, and increase the larger,
$\mu_2^{-1}$. Thus gradient coupling tends to increase the disparity in
these length scales.}

As in sections \ref{joco} and \ref{densdens} it is instructive to
see what happens to the parameters of the uncoupled model ($\nu=0$) as
$\nu>0$ is turned on. This can be extracted from the above formulae by
expanding in $\nu$, keeping only terms up to order $\nu^1$. 
One sees that
\bea
u_a(\nu)&=&\hat{u}_a\quad\mbox{exactly}\nonumber\\
\mu_A(\nu)^2&=&\hat\mu_A^2-2\nu \hat{u}_1\hat{u}_2\quad\mbox{exactly}\nonumber\\
\mu_a(\nu)^2&=&\hat\mu_a^2+O(\nu^2)
\eea
so, to leading order in $\nu$, the only length scale that changes is the 
penetration depth $1/\mu_A$, which increases.
The eigenvectors of
$\wt{\hh}$ are
\beq
v_1=\left(\begin{array}{c}1\\\frac{\hat{\mu}_1^2\nu}{\hat{\mu}_1^2-\hat{\mu}_2^2}\end{array}\right)+O(\nu^2),\quad
v_2=\left(\begin{array}{c}\frac{-\hat{\mu}_2^2\nu}{\hat{\mu}_1^2-\hat{\mu}_2^2}
\\1\end{array}\right)+O(\nu^2)
\eeq
which, one should note, are {\em not} orthogonal: the angle between them
is $\frac\pi2-\nu+O(\nu^2)$. It follows that the vortex has asymptotic
densities (at large $r$)
\bea
|\psi_1|&\sim&\hat{u}_1+q_1K_0(\hat\mu_1r)-\frac{q_2\hat\mu_2^2\nu}{\hat\mu_1^2-\hat\mu_2^2}K_0(\hat\mu_2r)+O(\nu^2)\nonumber\\
|\psi_2|&\sim&\hat{u}_2+q_2K_0(\hat\mu_2r)+\frac{q_1\hat\mu_1^2\nu}{\hat\mu_1^2-\hat\mu_2^2}K_0(\hat\mu_1r)+O(\nu^2)
\eea
where $q_1,q_2$ are unknown constants. So, while the ``coherence lengths'' 
remain unchanged to leading order, the normal modes with which they are
associated do receive a correction at order $\nu^1$. 

Finally, we remark that an alternative approach to handling gradient-gradient
terms is to remove them from $F$ from the outset by a linear redefinition
of the fields: essentially one expands $(\psi_1,\psi_2)^T$ in a basis of
eigenvectors of $P(\nu)$ \cite{notpublished}. This is mathematically elegant, but tends to
obscure the physical meaning of the non-gradient terms $F_p$.

\section{{Numerical solution of the nonlinear problem.}}
\label{numerics}

\begin{figure}
\begin{center}
\includegraphics[width=90mm]{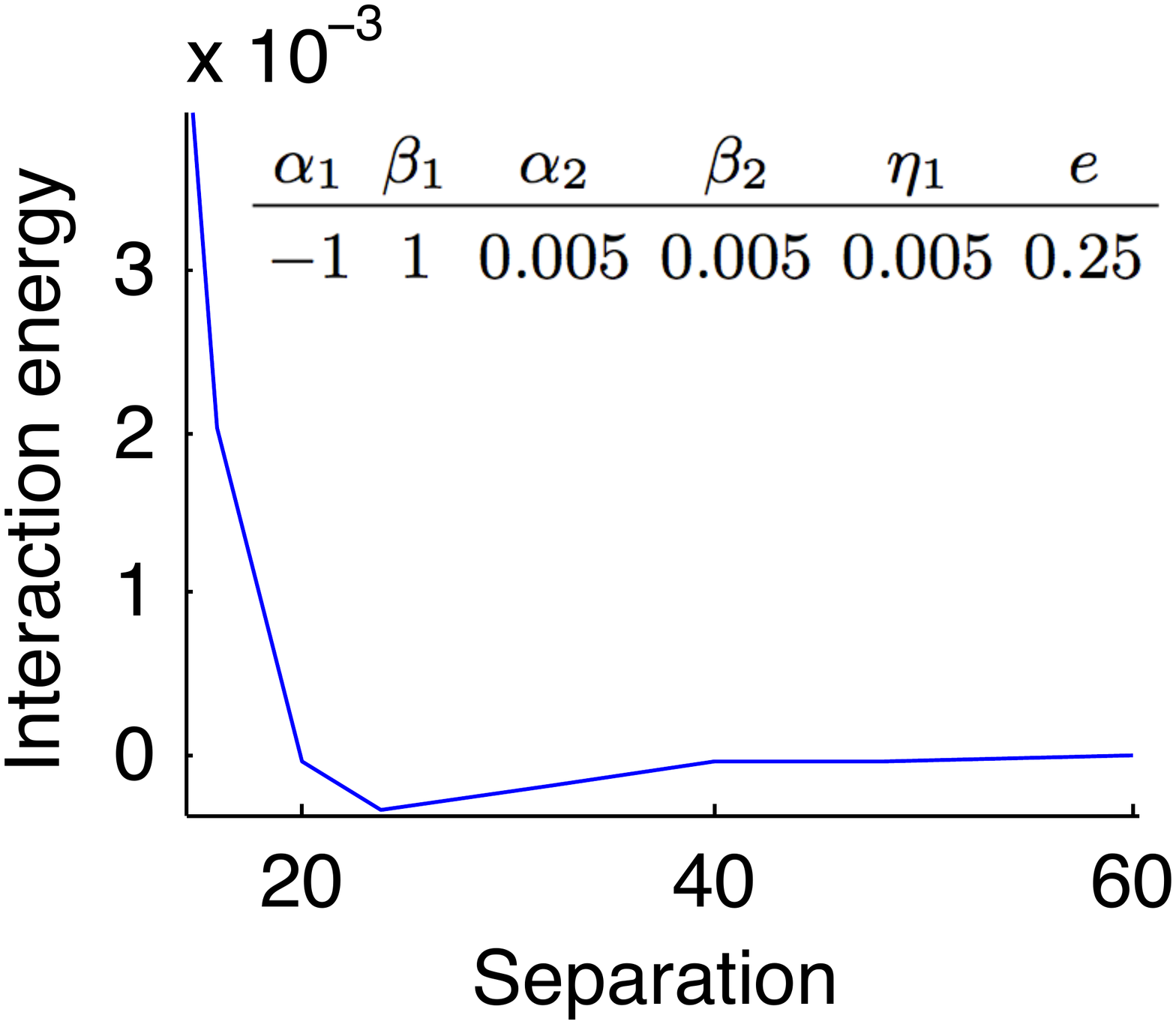}
\end{center}
\caption{Non-monotonic vortex interaction in a system with a passive band (i.e.
superconductivity is induced in the second band by an intrinsic  proximity effect).
In limit of zero Josephson coupling,  the active band would have $\kappa=8\kappa_c$, and thus would be deep into the type-II region.
This figure shows that a perturbation in the form of a weak Josephson coupling to passive band in this case produces a minimum
in the intervortex potential at a very large distance from the
vortex center.
}
\label{f1}
\end{figure}

\begin{figure}
\begin{center}
\includegraphics[width=90mm]{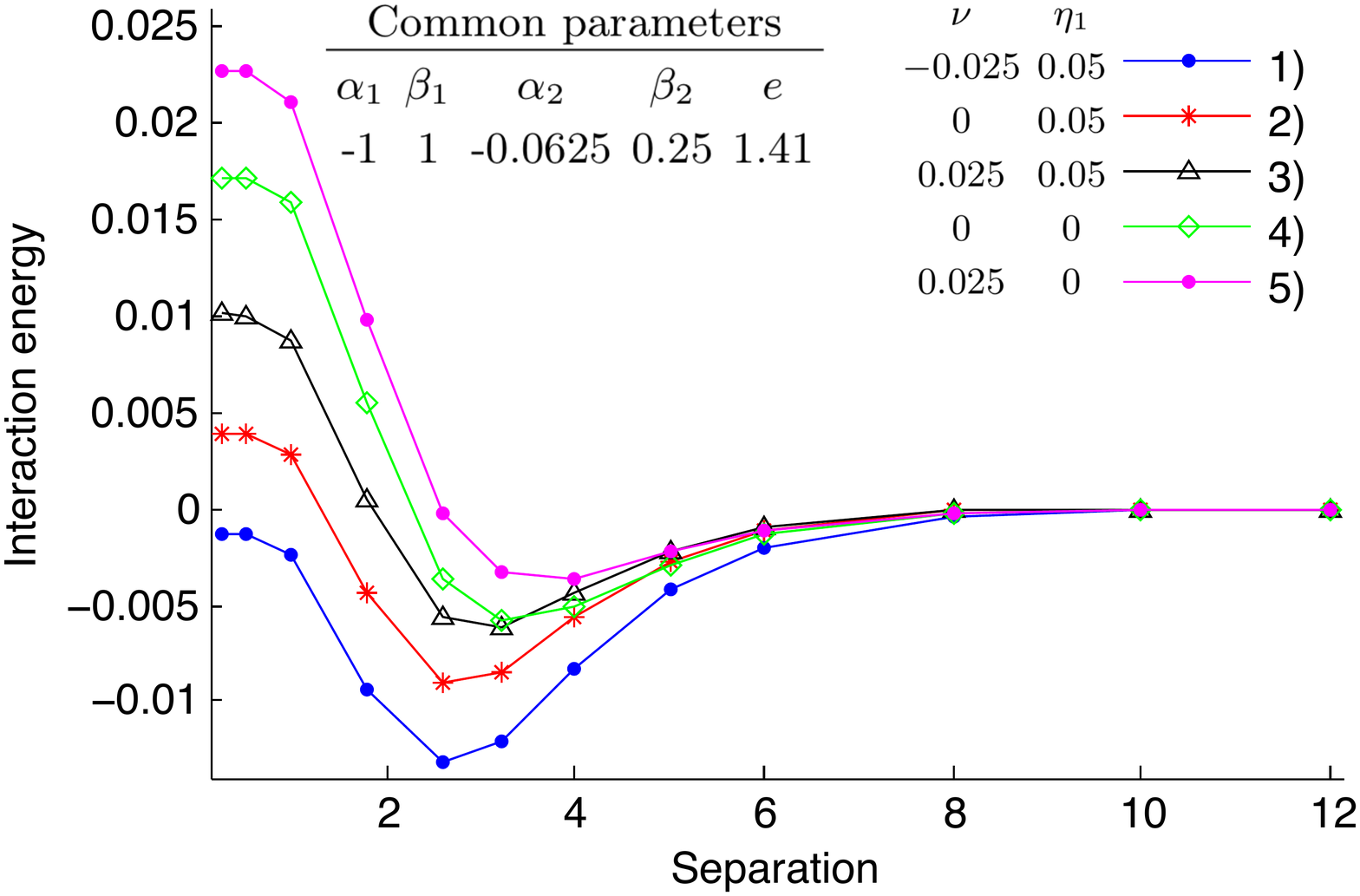}
\end{center}
\caption{Intervortex interaction potential
for a set of systems with two active bands. The systems share the parameters given in the table 'common parameters'.  
The green curve (4) corresponds to the case where the bands 
are coupled by the vector potential only. In this case, the 
ratio of the coherence lengths is $\xi_2/\xi_1=4$. 
The curve (2) shows the effect of the addition of Josephson term,
the curve (5) shows the effect of addition of mixed gradient term.
The curves (1) and (3) show the effect of the presence of
both mixed gradient and Josephson terms with similar
and opposite signs.
}
\label{f2}
\end{figure}

\begin{figure}
\begin{center}
\includegraphics[width=90mm]{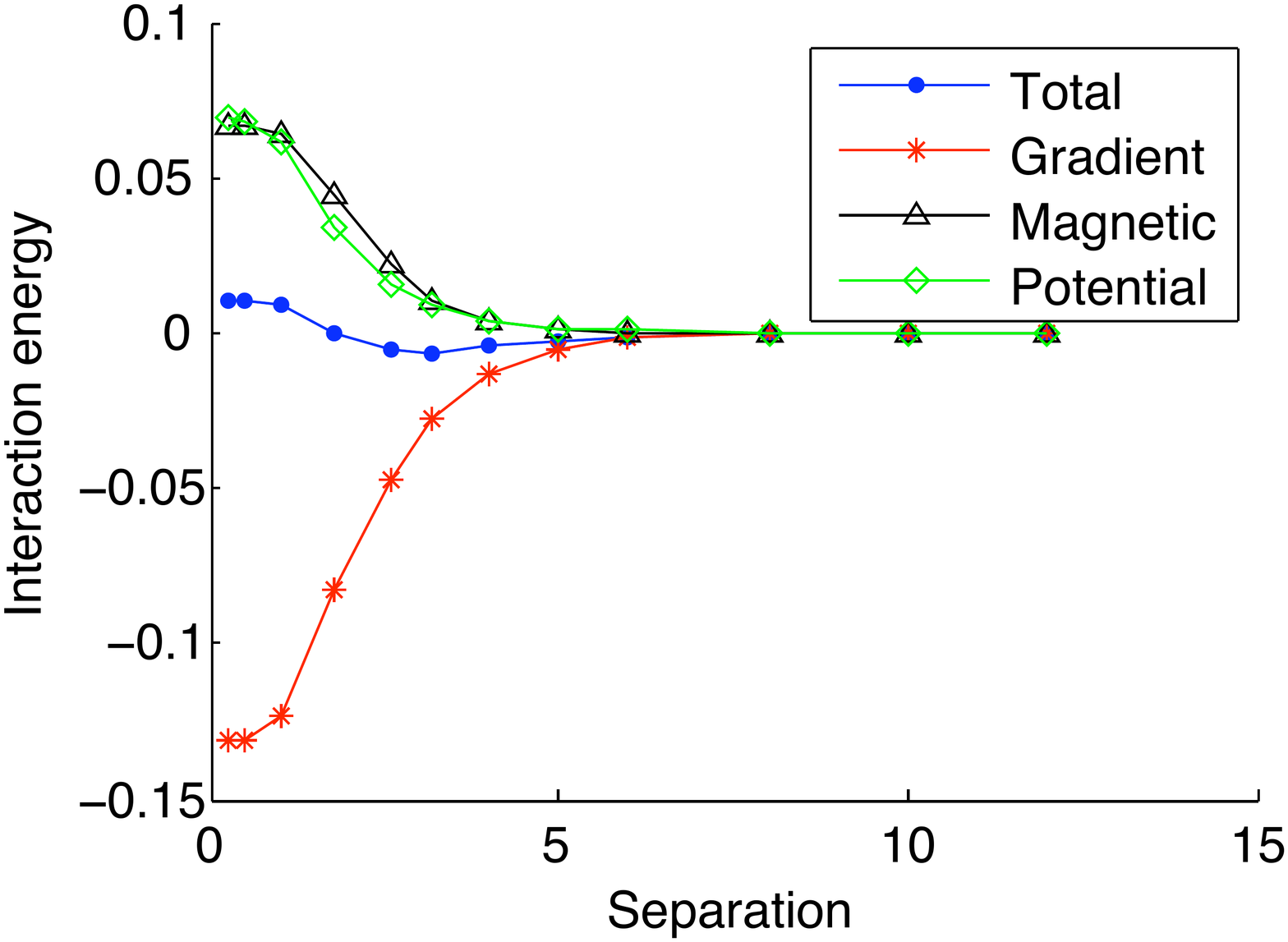}
\end{center}
\caption{Gradient, magnetic and potential energy 
contributions to the vortex interaction energy for the 
parameter set corresponding to the curve (3) of the Fig. \ref{f2} (black curve).  }
\label{f2b}
\end{figure}

\begin{figure}
\begin{center}
\includegraphics[width=140mm]{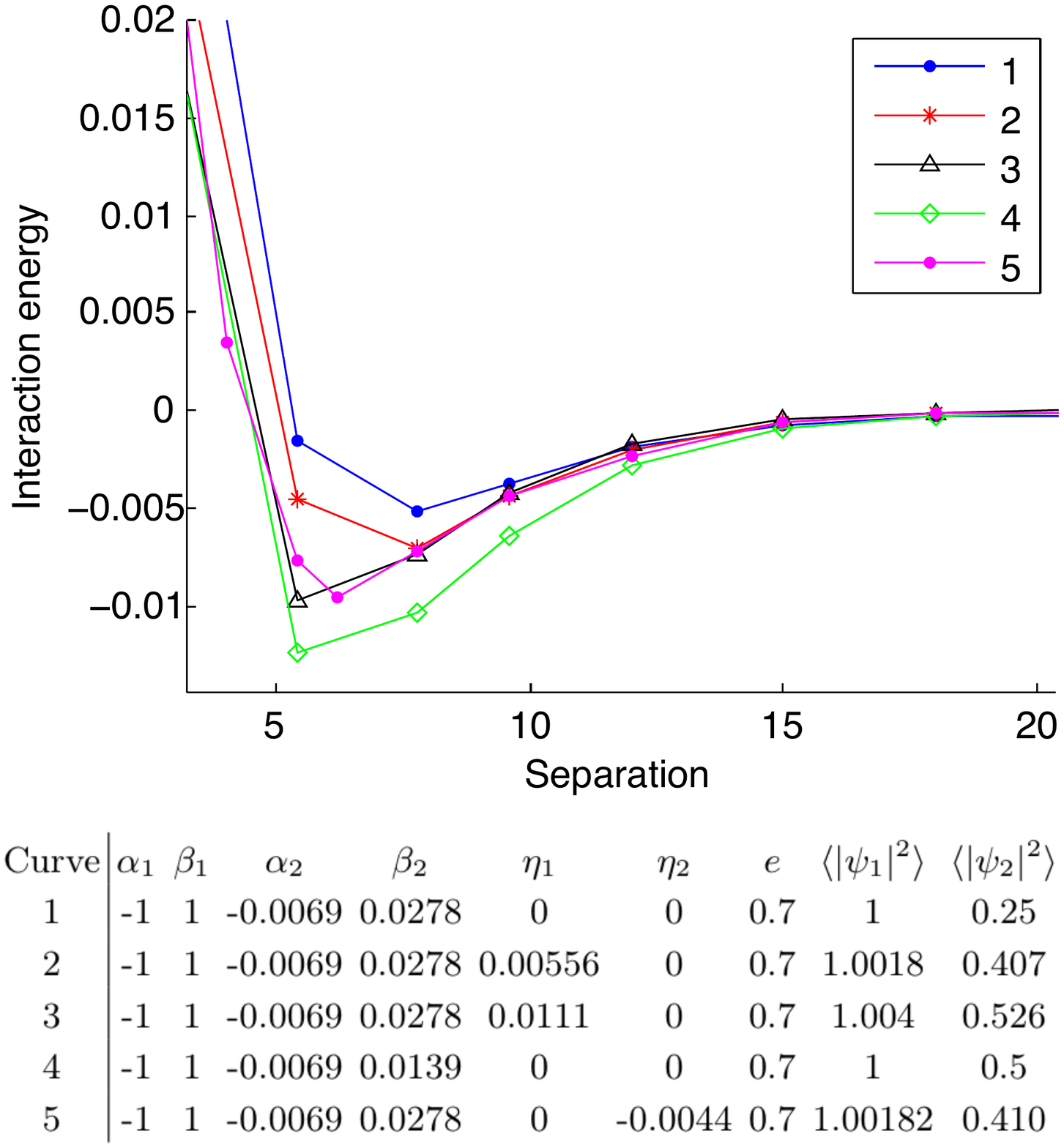}
\end{center}
\caption{Non monotonic vortex interaction in systems with two active bands and significant
disparity in length scales associated with the density variation.
 In the absence of inter band coupling (curve 1), the ratio of the coherence length is $\xi_2/\xi_1=12$. }
\label{f3}
\end{figure}

\begin{figure}
\begin{center}
\includegraphics[width=106mm]{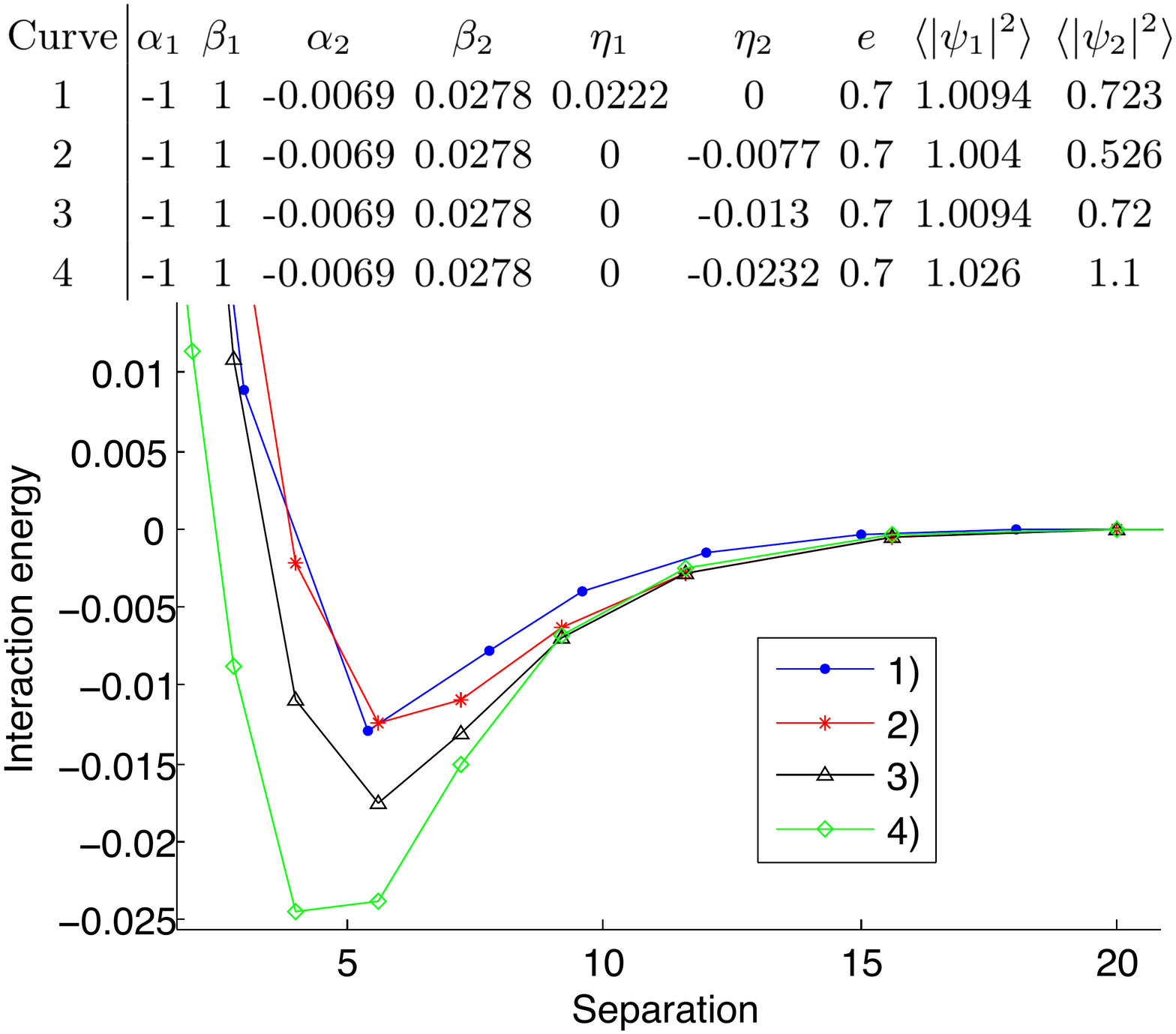}
\end{center}
\caption{Comparison of how Josephson coupling (1) and high order density coupling (2-4) affects vortex interaction energy. In (3), $\eta_2$ is chosen so that the densities are approximately the same as in (1). In (4), $\eta_2$ is chosen to give the same condensation energy as (1). Both these parameter values give larger vortex binding energy. Similar binding energy as (1) is acquired for a significantly smaller $\eta_2$ (2). The large condensation energy associated with Josephson coupling is responsible for the shorter interaction range in (1).
 }
\label{f3b}
\end{figure}


\begin{figure*}
\begin{center}
\includegraphics[width=200mm]{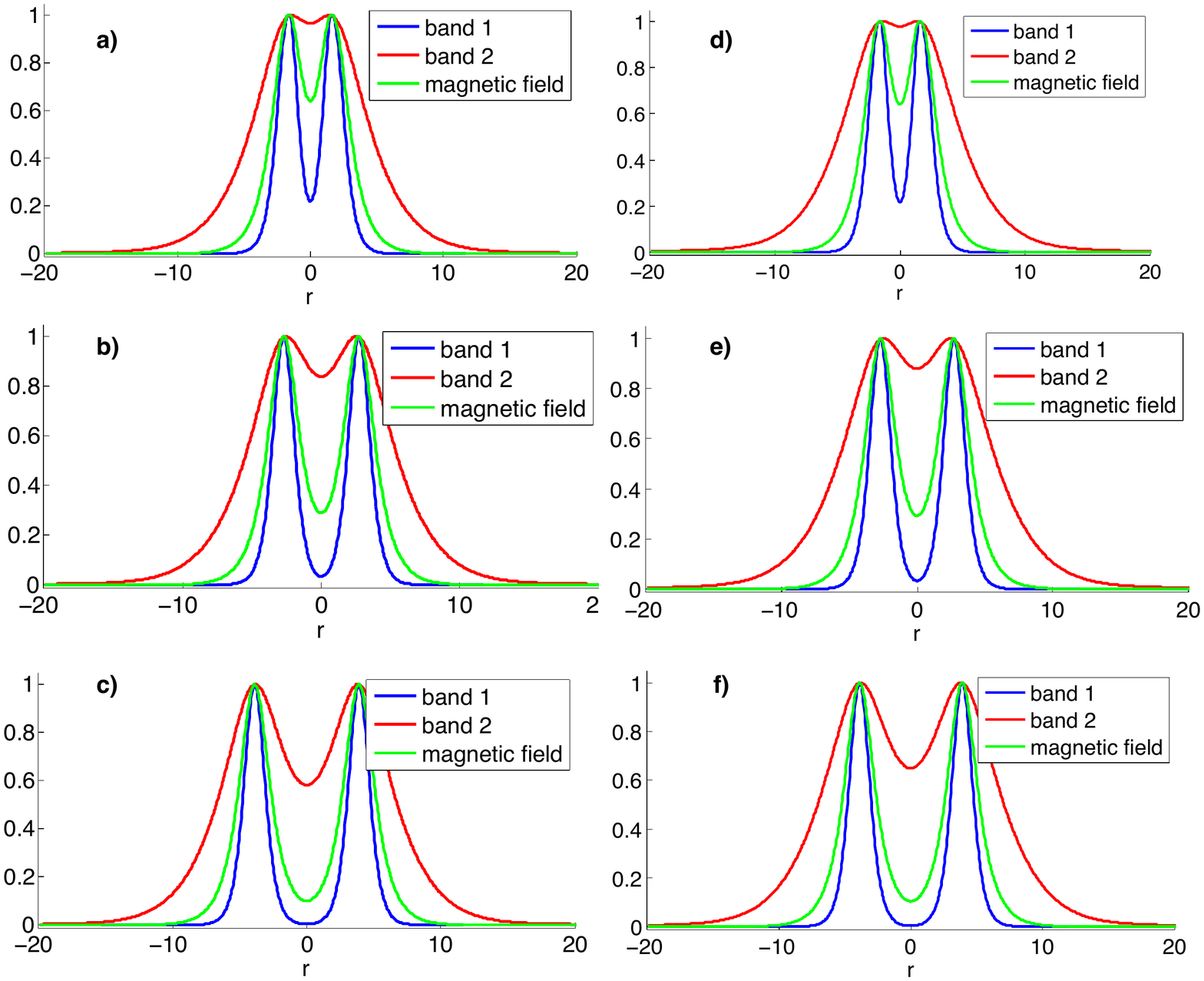}
\end{center}
\caption{Cross sections of two interacting vortices. The densities are given as $1-|\psi_i|^2/\langle |\psi|^2\rangle$, i.e. the condensates are normalized to $1$ and turned upside down. The magnetic field is also normalized. The left column corresponds to curve 3) of figure (\ref{f3}) and the right column correspond to curve 4) of the same figure. The inter-vortex separation is 3.0 in the upper row, giving repulsion, 5.4 in the mid row corresponding to the minimum energy, and 7.8 in the bottom row giving attraction. 
The curves 3 and 4 in fig \ref{f3} are the ones showing the largest binding energy, a property resulting from them having the largest vacuum expectation values in the second band ($0.526,\;0.5$). A careful comparison shows that the recovery of the second band is slower in the right column where there is zero inter band coupling. This is a very generic result, Josephson coupling shrinks the disparity in length scales.
The result of this effect is also seen when comparing the vortex interaction energies of the two system, the curve 3) reveals a shorter vortex interaction range than curve 4). }
\label{f4}
\end{figure*}

\begin{figure}
\begin{center}
\includegraphics[width=120mm]{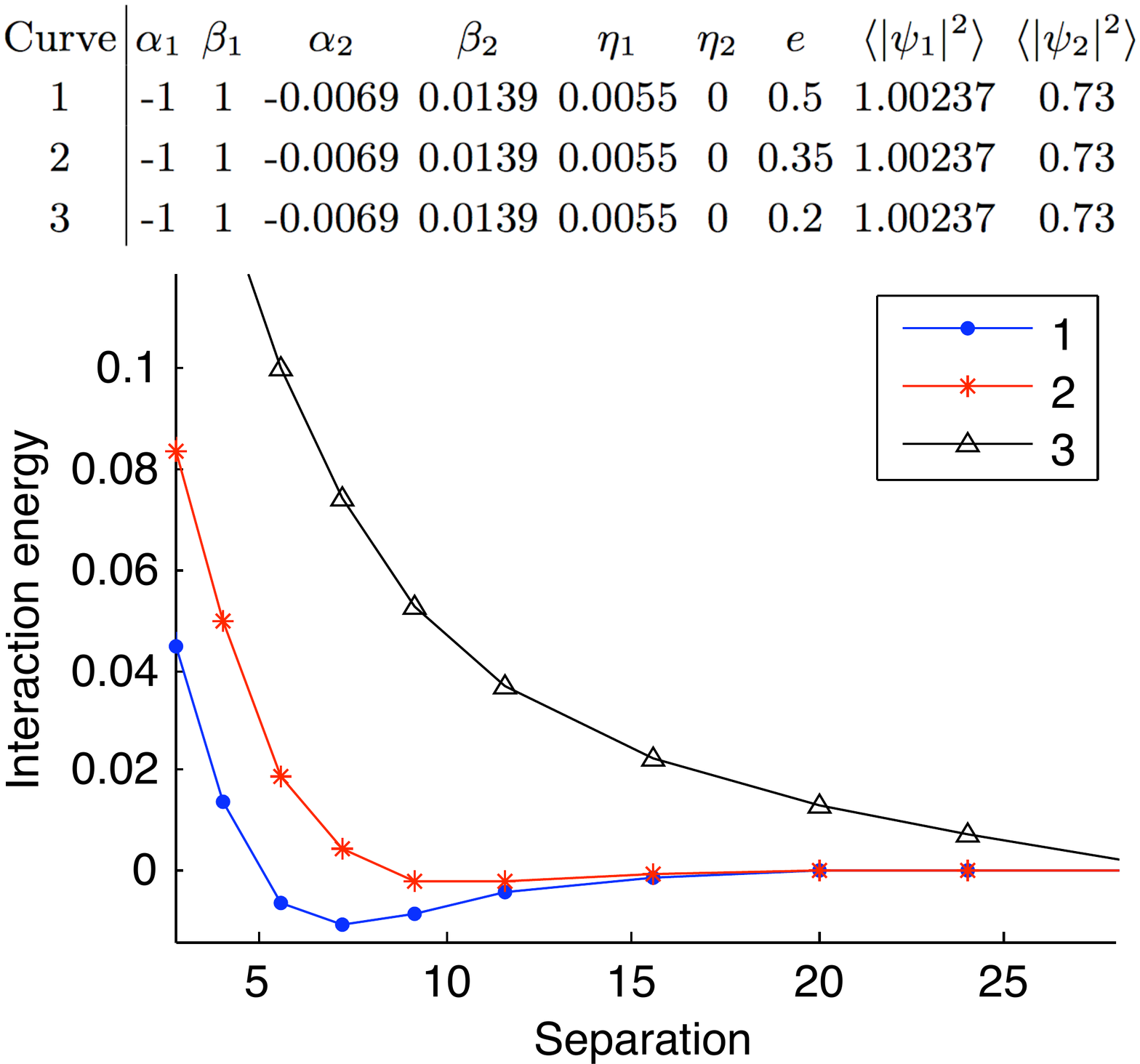}
\end{center}
\caption{Transition from type-1.5 to type-II region in a system with two active bands. The binding energies are $0.0109$ in the first curve and  $0.0023$ in the second curve. The third curve give monotonic repulsion. The small binding energy in the second case signals that
in this particular example the critical value for $e$ is close to $0.35$.  }
\label{f5}
\end{figure}
The linear analysis presented in the previous section can only provide information about the asymptotic tail
of the intervortex interaction. To determine the actual full intervortex
potential, especially in the case of strong interband coupling, it is necessary to treat the full nonlinear Ginzburg-Landau theory, something which is not possible analytically.
In this section we present interaction energies for vortex pairs (including cases of relatively strong interband coupling), computed
numerically  using a local relaxation method. The
numerical method which we use is the following: {A lattice approximant of the} energy is  minimized 
with respect all the degrees of freedom in the full Ginzburg-Landau functional
subject to the constraint that vortex positions remain fixed. 
This gives
 the intervortex interaction energy as a function of inter vortex separation. We used high resolutions grids, with the number of data points ranging from $1600\times1600$ to $2400\times1700$ and relaxed each configuration for $50-100$ hours on an eight core cluster node. 

{In this section 
the length scale is given in units of $2/\hat\mu_1$, where, as in section
\ref{asymp}, $\hat\mu_1$ denotes the mass 
of the field $|\psi_1|$ in the absence
of interband couping ($\eta_1=\eta_2=\nu=0$). Alternatively, the unit of
length is
$\sqrt{2}\hat{\xi}_1$, where $\hat\xi_1=\sqrt2/\hat\mu_1$ 
is the coherence length of the first band in the uncoupled case. 
Recall that $\hat{\xi}_1$ cannot be identified with physical coherence length
when interband coupling is present.
We also measure condensate density $|\psi_a|$ in units of $\hat{u}_1$, the 
vacuum expectation value of $|\psi_1|$ in the uncoupled case. As shown
in appendix \ref{appA}, this amounts to using scale freedom to
set $\alpha_1=-1$ and $\beta_1=1$. In the single component limit, the parameter $e$ can then be interpreted
as an inverse GL parameter.
 More precisely, $\kappa=\sqrt{2}/e$, so
in these units, in single-component limit, the critical value of $e$ is $e_c=2$.
The inter vortex interaction energy is given in units of total vortex energy, i.e.\ $2 E_v$ where $E_v$ is the energy of a single isolated vortex. All energies
are measured relative to the uniform Meissner state ($\psi_a=u_a$, $A=0$,
in the notation of section \ref{asymp}).}

\subsection{Weak Josephson coupling to a passive band}
Let us  consider a strongly type-II single-component superconductor, and see how vortex interaction in this system is modified by a weak Josephson coupling to a passive band 
(i.e. a band which 
has no superconductivity of its own which 
in the context of GL theory manifests itself as a positive coefficient $\alpha_2$).

The Fig. \ref{f1} shows the vortex interaction energy in such a system.
In the limit of decoupled bands, the parameter $e$ is here $e_c/8$. 
Therefore, for zero Josephson coupling this system would have $\kappa\approx 5.7$, putting it far into the type-II region. 
Weak Josephson coupling changes the length scales as
discussed in the previous section and adds a qualitatively new feature: the inter vortex
potential acquires a minimum, 
occurring at a separation of  approximately $24\sqrt{2}\xi_1$.

\subsection{Effects of Josephson and mixed gradient terms in case of two active bands}
The figure \ref{f2} illustrates the effect of the mixed gradient term, as well as of Josephson coupling in a system with two active bands. The green curve (4) corresponds to two independent bands, interacting only trough the magnetic field. The curve (2) corresponds to a system with added Josephson coupling which increases the binding energy, but decreases the 
distance where the energy minimum is located and slightly reduces the range of interaction. 

The inclusion of a mixed gradient term (shown as curve (5)) here has a similar effect on phase difference as the Josephson term. When the phases are locked $\theta_1-\theta_2=0$, effectively this term gives
 a negative contribution to the energy associated with co-directed currents. Thus, for this choice of the sign of {$\nu$}, the mixed gradient term also prefers phase locking $\theta_1-\theta_2=0$.

Due to symmetry, changing $\eta_1\to-\eta_1$ and $\nu\to-\nu$ does not qualitatively change the behavior of the system, as this only results in phase locking with $\pi$ difference instead. 
While Josephson coupling increases the energy of vortices, and mixed gradients decreases it, their effect on interaction energy is the opposite. The decomposition of vortex interaction energy into a set of contributions
from different terms given in Fig \ref{f2b} illustrates why mixed gradients in this case increase repulsion.

In contrast the curve (1) (blue curve), corresponds to the case where $\nu$ and $\eta_1$ have different signs, and so there is competition between the gradient mixing and the Josephson term with regard to 
the preferred phase difference.
The mixed gradient term is minimal for a phase locking where $\theta_1-\theta_2=\pi$, while the Josephson term is minimal for $\theta_1-\theta_2=0$. 
The result in these simulations was that the phase locking was determined by the dominating Josephson coupling, and that the gradient mixing resulted in increased cost for co-directed currents. This was the most energetically expensive 
vortex, but also exhibited the smallest inter vortex interaction energy.

\subsection{Solutions with large disparity in the characteristic length scales}

Figure \ref{f3} shows a set of simulations done with two active bands and a larger disparity in characteristic length scales. 
We start 
with case when the condensates interact only through the magnetic field (blue curve (1)), and the density in the second band is $1/4$ of the density in the first band and the coherence length ratio being $\xi_2/\xi_1=12$ {(so in the notation of
section \ref{asymp}, $\hat{u}_1=4\hat{u}_2$ and $\hat\mu_1=12\hat\mu_2$)}.
This allows non-monotonic interaction to occur at smaller $e$ than above - here we simulate at $e=0.7$. This gives the smallest binding energy in this set of simulations. Adding a Josephson coupling of $\eta_1=0.00556$ [shown on curve (2)] gives a substantially higher density in the second band, and 
thus  a stronger binding energy. Adding a stronger Josephson coupling [shown on the curve (3)] gives even larger density in the second band. It also shows some of the qualitative differences associated with this coupling. Namely, the binding occurs at a smaller separation, and the range of the interaction decreases. It is clearly visible that  curve (3) crosses the other curves. The Josephson coupling causes the second
condensates to recover faster (as follows also from the linear theory 
presented in the previous section), and thereby decreasing the range of attractive interaction. 

Decreasing $\beta_2$ raises the density of condensate in the second band. In curve (4), we have reduced it by a factor 2, thereby increasing the vacuum expectation value of the density in the second band by a factor 2. This does however not change the length scale, as $\xi$ is independent of $\beta$, but it does increase the energetic benefits of core overlap in the second component. This case  shows the largest binding energy in this set of simulations.

Finally, we consider the effect of a higher order density-density coupling $\eta_2$ between the condensates which is shown on the curve (5). The parameter choice here gives approximately the same densities as (2) but smaller condensation energy.
This should generally make the system (5) recover slower than system(2), resulting in longer range of the interaction. 
{ A more systematic comparison of Josephson coupling and higher order density coupling supporting this conclusion is given in }Fig. \ref{f3b}.

Figure \ref{f4} displays cross sections of the condensate densities and magnetic flux in the systems 3-4 in Fig. \ref{f3}, clearly illustrating the mechanism by which type-1.5 superconductivity appears.

Let us consider a different example of the appearance of the type-1.5 regime in the case where there is a substantial disparity in the dominant length scales associated
with the variations of densities and magnetic field penetration length.

Again, our starting point is a reference case of
two bands coupled only by vector potential
where we choose $\alpha_2$ so that the disparity in coherence length in absence of inter band coupling is $\xi_2/\xi_1=12$. 
\begin{figure}
\begin{center}
\includegraphics[width=70mm]{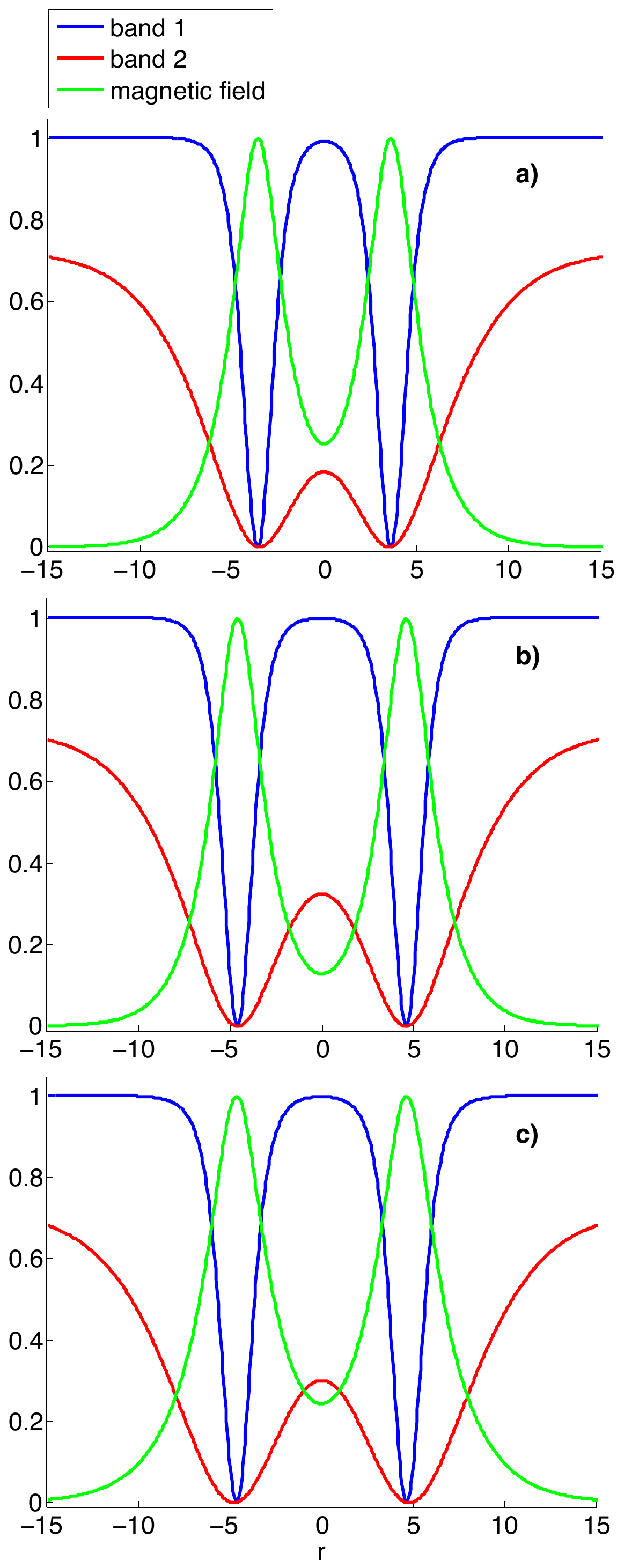}
\end{center}
\caption{Cross sections plots showing condensate density and magnetic flux in type-1.5 superconductors with small $e$. a) Curve (1) in Fig. \ref{f5} at a separation of $7.2$, corresponding to the energy minimum. b) The same system at a separation of $9.6$. c) Curve (2) of the same figure at a separation of $9.6$, corresponding to the energy minimum. The three distinct length scales associated with two band superconductors are indeed visible. Despite the inter-band Josephson coupling, there is a significant disparity in the recovery of the two condensates in all the plots. The third length scale, penetration depth, visibly differs between the two systems, and is responsible for the differences in inter vortex interaction.}
\label{f5b}
\end{figure}

Now, we take $\beta_2$ to be $0.0139$ i.e. the same as in curve (4) of Fig. \ref{f3} and choose the Josephson coupling to be $\eta_1=0.00556$. Then, we successively decrease $e$ to see where the system crosses over from type-1.5 to type-II. The outcome can be seen in Fig. \ref{f5}. The first curve gives a binding energy of $0.011$, the second gives $0.0023$ and the third curve shows system 
crossing over into type-II regime by showing  monotonic repulsion. Given the small binding energy of the second curve, the cross over from type-1.5 to type-II, in this example, is close to $e=0.35$. 

The crossection plots of two of these systems given in Fig \ref{f5b} illustrate how the system crosses over from type-1.5 to type-II as $e$ is decreased
resulting in dominance of the repulsive force originating in the  electromagnetic and current-current interaction.

\section{Conclusions}

In this paper we presented an analytic and numerical study
of the appearance
of type-1.5 superconductivity in the case of two bands with various kinds of
substantial interband couplings.
In all the cases which we considered we demonstrated that the system
possessed three fundamental length scales:
one length scale $1/\mu_A$ associated with London magnetic field
penetration length while the
other two fundamental scales $1/\mu_{1,2}$ are associated with
characteristic length scales controlling variations of density fields.
In the limit of two condensates coupled only electromagnetically
the length scales $1/\mu_{1,2}$ are associated with independent
coherence lengths of two fields.
However we show that introducing a nonzero Josephson and quartic
density-density
couplings makes both density fields
decay according to the same exponential law at very large distances
from the core
while, {\it at the same time the system still possesses two fundamental
length scales
which are associated now with
variation of linear combinations  of density fields rotated by a
``mixing angle"}.
The third fundamental length scale in that regime is the London
penetration length and thus
two-band systems with  
couplings allow a well defined type-1.5 behavior.
Next we studied the effect of mixed gradient terms
and showed how the type-1.5 regime is described in that case.
We showed that 
in the case of a substantial mixed gradient coupling the
definition of three  fundamental
length scales requires   additional care because
it produces mode mixing which cannot be described by a single
mixing angle. 
Importantly, we demonstrated that 
mixed gradient coupling can enhance the disparity
of the characteristic length scales of the density variations.
An analogy can be drawn between this mechanism and the Seesaw mechanism in neutrino physics.
In the second part of the paper we presented
a comparative numerical study of
type-1.5 vortices in the different regimes with various intercomponent
couplings.
The results were demonstrated in the framework of  a two-component 
Ginzburg-Landau model
with local electrodynamics.
However we expect that the described type-1.5 behavior
is similarly present in lower-temperature regimes and in
two-component models with non-local electrodynamics.

The concept of type-1.5 superconductivity can be straightforwardly generalized to N-component case. There it can occur in systems where characteristic length scales are $\xi_1,..,\xi_k<\lambda<\xi_{k+1},...,\xi_N$ and there are thermodynamically stable vortices with nonmonotonic interaction.
 

Besides multi-band superconductors and coexistent electronic and nuclear superconductors our model can be realized in 
artificially fabricated layered structures made of type-II and type-I
materials where one can control and tune intercomponent 
Josephson coupling.

{{\it Note added:} After the completion of this work,  it was  verified in a  microscopic
calculation which does not involve  $(1-T/T_c)$ expansion, that the GL model (\ref{gl}) should quite accurately describe the vortex 
physics in two-band superconductors in a quite wide range of parameters and temperatures
\cite{silaev}.}

\section*{Acknowledgements}
We thank Alex Gurevich for discussions.
EB was supported by Knut and Alice Wallenberg
Foundation through the Royal Swedish Academy of Sciences, Swedish Research Council and by the US National
Science Foundation CAREER Award No. DMR-0955902.
JC was supported by the Swedish Research Council.
MS was supported by the UK Engineering and Physical Sciences
Research Council.


\appendix
\section{Units}\label{appA}
In this section we give an explicit mapping from our representation of the GL 
model to
a more common textbook representation.
Consider the Ginzburg-Landau model in the following quite usual units
\begin{eqnarray}
\nonumber
F&=&\frac{\hbar^2}{2m_1}\Big|\Big(\nabla-i\frac{e^*}{\hbar c}A\Big)\psi_1\Big|^2 \\\nonumber
&&+\frac{\hbar^2}{2m_2}\Big|\Big(\nabla-i\frac{e^*}{\hbar c}A\Big)\psi_2\Big|^2+ \\\nonumber
&&-\nu \hbar^2 Re\Big\{(\nabla-i\frac{e^*}{\hbar c}A)\psi_1\cdot(\nabla+i\frac{e^*}{ \hbar c}A)\psi_2^*\Big\}\\\nonumber
&&+\frac{1}{8\pi}(\nabla\times A)^2\\\nonumber
&&+\alpha_1|\psi_1|^2+\frac{1}{2}\beta_1|\psi_1|^4+\alpha_2|\psi_2|^2+\frac{1}{2}\beta_2|\psi_2|^4\\
&&-\eta_1|\psi_1||\psi_2|\cos(\theta_1-\theta_2)+\eta_2|\psi_1|^2|\psi_2|^2
\end{eqnarray}
Let us define the rescaled quantities
\bea
\wt{F}&=&\frac{4\pi}{\hbar^2 c^2}F\nonumber\\
\wt{A}&=&-\frac{A}{\hbar c}\nonumber\\
\wt{\psi}_a&=&\sqrt{\frac{4\pi}{m_ac^2}}\psi_a\nonumber\\
\wt{\nu}&=&\sqrt{m_1m_2}\nu\nonumber\\
\wt{\alpha}_a&=&\frac{m_a}{\hbar^2}\alpha_a\nonumber\\
\wt{\beta}_a&=&\frac{m_a^2c^2}{4\pi\hbar^2}\beta_a\nonumber\\
\wt{\eta}_1&=&\frac{\sqrt{m_1m_2}}{\hbar^2}\eta_1\nonumber\\
\wt{\eta}_2&=&\frac{m_1m_2c^2}{4\pi\hbar^2}\eta_2.
\eea
Then
\bea
\wt{F}&=&\frac12\left|(\nabla+ie^*\wt{A})\wt{\psi}_1\right|^2+
\frac12\left|(\nabla+ie^*\wt{A})\wt{\psi}_2\right|^2\nonumber \\
&&-\wt{\nu}{\rm Re}\left\{(\nabla+ie^*\wt{A})\wt{\psi}_1\cdot
(\nabla-ie^*\wt{A})\wt{\psi}_2^*\right\}\nonumber \\
&&+\frac12|\nabla\times\wt{A}|^2\nonumber \\
&&-\wt\alpha_1|\wt{\psi}_1|^2+\frac{\wt{\beta}_1}{2}|\wt{\psi}_1|^2
+\wt\alpha_2|\wt{\psi}_2|^2+\frac{\wt{\beta}_2}{2}|\wt{\psi}_2|^2\nonumber \\
&&+\wt\eta_1|\wt\psi_1||\wt\psi_2|\cos(\theta_1-\theta_2)
+\wt\eta_2|\wt\psi_1|^2|\wt\psi_2|^2,
\eea
which, on dropping the tildes, coincides with the representation 
(\ref{gl}) used in this paper.

Throughout the paper, it is assumed  that
band 1 is active, that is, $\alpha_1<0$. It
is convenient to rescale the expression (\ref{gl})
for $F$ further so that
$\alpha_1$ is normalized to $-1$ and $\beta_1$ is normalized to $1$.
This can be achieved as follows. Recall (see section \ref{basic}) that
in the absence of interband couplings (i.e.\ when $\eta_1=\eta_2=\nu=0$)
condensate 1 has decay length-scale $1/\hat{\mu}_1=(-4\alpha_1)^{-1/2}$. This
scale is more usually specified by the coherence length
\beq
\hat\xi_1=\frac{\sqrt{2}}{\hat{\mu}_1}=\frac{1}{\sqrt{-2\alpha_1}}.
\eeq
We emphasize once more that, in the presence of interband couplings,
$\hat\xi_1$ in not the coherence length of 
condensate 1.\,
This is the purpose of the hat, to remind us that this is a genuine
coherence length only in the uncoupled case. Recall also that the
vacuum density of condensate 1 in the uncoupled model is
\beq
\hat{u}_1=\sqrt{\frac{-\alpha_1}{\beta_1}}.
\eeq
Our second rescaling amounts to using $\sqrt{2}\hat\xi_1$ as the unit of
length and $\hat{u}_1$ as the unit of condensate density (along with
compensating rescalings of $F$, $e^*$ and $A$). Explicitly, let
\bea
\bar{r}&=&\frac{r}{\sqrt{2}\hat\xi_1}=\sqrt{-\alpha_1}r\nonumber\\
\bar{F}&=&\frac{2\hat\xi_1^2}{\hat{u}_1^4}F=\frac{\beta_1^2}{-\alpha_1^3}F
\nonumber\\
\bar{\psi}_a&=&\frac{\psi_a}{\hat{u}_1}=\sqrt{\frac{\beta_1}{-\alpha_1}}\psi_a
\nonumber\\
\bar{A}&=&\frac{A}{\hat{u}_1}\nonumber\\
\bar{e}&=&\frac{1}{\sqrt2}\hat{u}_1\hat\xi_1e^*=\frac{e^*}{\sqrt{\beta_1}}
\nonumber\\
\bar{\alpha}_2&=&2\hat\xi_1^2\alpha_2=\frac{\alpha_2}{-\alpha_1}
\nonumber\\
\bar{\beta}_2&=&2\hat\xi_1^2\hat{u}_1^2\beta_2=\frac{\beta_2}{\beta_1}
\nonumber\\
\bar{\eta}_1&=&2\hat\xi_1^2\eta_1=\frac{\eta_1}{-\alpha_1}
\nonumber\\
\bar{\eta}_2&=&2\hat\xi_1^2\hat{u}_1^2\eta_2=\frac{\eta_2}{\beta_1}
\nonumber\\
\bar{\nu}&=&\nu.
\eea
Substituting these into (\ref{gl}) yields
\bea
\bar{F}&=&\frac12\left|(\bar\nabla+i\bar{e}\bar{A})\bar{\psi}_1\right|^2+
\frac12\left|(\bar\nabla+i\bar{e}\bar{A})\bar{\psi}_2\right|^2\nonumber \\
&&-\bar{\nu}{\rm Re}\left\{(\bar{\nabla}+i\bar{e}\bar{A})\bar{\psi}_1\cdot
(\bar{\nabla}-i\bar{e}\bar{A})\bar{\psi}_2^*\right\}\nonumber \\
&&+\frac12|\bar\nabla\times\bar{A}|^2\nonumber \\
&&-|\bar{\psi}_1|^2+\frac{1}{2}|\bar{\psi}_1|^2
-\bar\alpha_2|\bar{\psi}_2|^2+\frac{\bar{\beta}_2}{2}|\bar{\psi}_2|^2\nonumber \\
&&-\bar\eta_1|\bar\psi_1||\bar\psi_2|\cos(\theta_1-\theta_2)
+\bar\eta_2|\bar\psi_1|^2|\bar\psi_2|^2.
\label{maya}
\eea
This (having dropped the bars)
is the GL energy used in section \ref{numerics} for the purposes of
numerical simulation.

Finally, we note that the single component GL model obtained from (\ref{maya})
by setting $\psi_2\equiv 0$ has penetration depth $\lambda=1/\mu_a=1/e$ and
coherence length $\xi=1/\sqrt{2}$, and hence GL parameter
\beq
\kappa=\lambda/\xi=\frac{\sqrt{2}}{e}.
\eeq
So, in the parameterization used in section \ref{numerics}, one may regard $e$
as an inverse GL parameter for the associated single band model. 
The value of $e$ corresponding to the Bogomolny limit of
one-component theory is $e_c=2$ in this interpretation.

\section{The spectrum of $\wt{\hh}$}
\label{appB}
\news

Let $\hh$ be a real, symmetric $2\times 2$ matrix both of whose
eigenvalues are positive, and let $\wt\hh(\nu)=P(\nu)^{-1}\hh$ where
$P(\nu)$ is defined in equation (\ref{gimo}). We wish to show that
the eigenvalues $\lambda_1(\nu),\lambda_2(\nu)$ of
$\wt\hh(\nu)$ are also real and positive
for all $0\leq\nu<1$. First note that $\wt\hh(0)=\hh$, so the conclusion
holds at $\nu=0$. Further, $\lambda_a(\nu)$ depend continuously on $\nu$.
Now $\det\wt\hh(\nu)=\det\hh/(1-\nu^2)\neq 0$, so neither eigenvalue ever
vanishes. So $\lambda_a(\nu)$ remain real and positive, unless, for some
$\nu=\nu_*\in(0,1)$, they coalesce and bifurcate into a complex conjugate
pair. But then, at $\nu=\nu_*$ we have $\lambda_1(\nu_*)=
\lambda_2(\nu_*)=\lambda_*\in(0,\infty)$ and hence 
$\wt{\hh}(\nu_*)=\lambda_*\I_2$.
But then 
\beq
\wt{\hh}(\nu)=P(\nu)^{-1}P(\nu_*)\wt{\hh}(\nu_*)=\lambda_*P(\nu)^{-1}P(\nu_*)
\eeq
which is symmetric,  and hence has real eigenvalues, for all $\nu\in(0,1)$.
Hence a bifurcation to a complex conjugate pair of eigenvalues is not
possible, and we conclude that $\wt{\hh}(\nu)$ has real, positive eigenvalues
for all $\nu\in[0,1)$.

\section{Calculation of long range inter-vortex forces from the linear field asymptotics.}
\label{appC}
\news
{Here we outline how asymptotic intervortex forces are calculated from the linearized
asymptotic behavior  of the fields.  In the above we use a two-component generalization of the method
previously developed by one of us in the context of the single component GL model \cite{spe}.}
The key idea is to identify the vortices with static {\em topological solitons}
in a relativistic extension of the TCGL model to $2+1$ dimensional Minkowski space, which could be called a
two component abelian Higgs model. Viewed from afar, the solitons in this theory are identical to the fields
induced by suitable point sources in the linearization of the model, so the forces between well-separated solitons
approach those between the corresponding { fictitious  point particles, mediated by linear fields. The nature (attractive or
repulsive) and range of such forces can then be  computed}. 
In this appendix we take the opportunity
to explain the method in the simplest possible context, with the aim of making it more transparent to a
wide readership. Despite being simple and pedagogically motivated, the calculation we present is,
as far as we are aware, new, although the result itself has been derived previously by other means.

Consider the sine-Gordon model, which consists of a single scalar field $\phi$ 
in $1+1$ dimensional Minkowski space, evolving according to the Euler-Lagrange equation for the
action $S=\int\Ll dt\, dx$ with Lagrangian density
\beq
\Ll=\frac12\cd_\mu\phi\cd^\mu\phi-(1-\cos\phi).
\eeq
It is useful to think of $\phi$ as an angular variable, with period $2\pi$, so that the model has a single
ground state (or ``vacuum"), $\phi=0$. It also has static kink $\phi_K$ solutions in which $\phi$ tends
to the vacuum as $x\ra\pm\infty$ but winds once 
 anti-clockwise. Explicitly,
\beq
\phi_K(x)=4\tan^{-1}e^{x-x_0},
\eeq
where $x_0$ is a free parameter.
These are topological solitons (topologically stable, spatially localized lumps of energy) analogous to the
vortices of the GL model. Their energy density ${\cal E}=\frac12( \frac{\partial\phi}{\partial x})^2+(1-\cos\phi)$
is localized in a lump centred at $x=x_0$. At large $|x|$ the kink has asymptotic form
\beq\label{aski}
\phi_K(x)\sim\left\{\begin{array}{cc}
4e^{-|x-x_0|}&x\ra-\infty\\
2\pi-4e^{-|x-x_0|}&x\ra\infty.
\end{array}\right.
\eeq
We wish to identify this with the field induced by a suitable point source in the linearization of the field theory 
about the ground state $\phi=0\equiv 2\pi$. The linearized field theory has Lagrangian density
\beq
\Ll_0=\frac12\cd_\mu\phi\cd^\mu\phi-\frac12\phi^2
\eeq
obtained by expanding $\Ll$ about $\phi=0$ to quadratic order. The corresponding Euler-Lagrange equation is the Klein-Gordon
equation $\phi_{tt}-\phi_{xx}+\phi=0$ for a scalar field of mass $1$,
whose general static solution is
\beq
\phi=c_1e^{x}+c_2 e^{-x}.
\eeq
None of these reproduces the asymptotic kink on the whole real line: one needs $c_2=0$ for $x<x_0$ and $c_1=0$ for
$x>x_0$. To reproduce the kink's asymptotics we must introduce a source term into $\Ll_0$,
\beq
\Ll_0\mapsto\Ll_0+\kappa\phi
\eeq
where $\kappa(x)$ is the source. The field equation is now
\beq\label{emblu}
\phi_{tt}-\phi_{xx}+\phi=\kappa(x).
\eeq
If we take $\kappa$ to be a scalar monopole of charge $q$ placed at $x=0$, that is, $\kappa(x)=q\delta(x)$, 
the induced field is $\phi(x)=\frac{q}{2}e^{-|x|}$. We deduce that the asymptotic kink (\ref{aski}) will be 
induced by the point source 
\beq
\kappa_K(x)=m\delta'(x-x_0),\qquad m=8
\eeq
which may be interpreted as a scalar dipole of moment $m$ (here $\delta'$ denotes the derivative of the delta
distribution). 

So, viewed from afar, the kink soliton is identical to a scalar dipole in the linear theory. On physical
grounds, the interaction energy of a pair of kinks held a fixed distance apart should therefore approach  that between
a pair of scalar dipoles as the distance grows large. The latter interaction energy is easily computed.
Consider the field $\phi$ induced by a static source $\kappa(x)$ which is itself the sum of two static sources 
$\kappa_1(x)+\kappa_2(x)$. Since the field theory is linear, $\phi=\phi_1+\phi_2$, where $\phi_i$ is the field
induced by $\kappa_i$. The total action of the field $\phi$ and source $\kappa$ is
\bea
S&=&\int\bigg(\frac12\cd_\mu(\phi_1+\phi_2)\cd^\mu(\phi_1+\phi_2)-\frac12(\phi_1+\phi_2)^2\nonumber\\
&&+(\kappa_1+\kappa_2)(\phi_1
+\phi_2)\bigg)dx\, dt\nonumber\\
&=&S_1+S_2+\int\kappa_1\phi_2\, dx\, dt
\eea
where $S_i$ is the action of $(\phi_i,\kappa_i)$, and we have integrated by parts and used the fact that $\phi_2$ satisfies (\ref{emblu}) with source $\kappa_2$. From this we extract the interaction Lagrangian for the pair of sources $\kappa_1,\kappa_2$,
\beq
L_{int}=\int\kappa_1\phi_2\, dx.
\eeq
The case of interest is where $\kappa_i$ are scalar dipole sources of moment $m_i$ located
at $x_i$. One then has $\kappa_1(x)=m_1\delta'(x-x_1)$ and $\phi_2(x)=\frac12 m_2 \frac{d\:}{dx}e^{-|x-x_2|}$, 
and so
\bea
L_{int}&=&\int m_1\delta'(x-x_1)\phi_2(x)\, dx\nonumber\\
&=&-m_1\left.\frac{d\psi_2}{dx}\right|_{x=x_1}\nonumber\\
&=&-\frac12m_1m_2e^{-|x_1-x_2|}.
\eea
Since the interaction Lagrangian depends only on the dipoles' positions, we can identify $-L_{int}$ as the
interaction {potential},
\beq\label{isfi}
V_{int}=\frac12m_1m_2e^{-|x_1-x_2|}.
\eeq
From this we see that like scalar dipoles ($m_1,m_2$ same sign) repel one another, while unlike dipoles ($m_1,m_2$ 
opposite signs) attract. In the case of kinks, $m_1=m_2=8$, so
\beq
V_{KK}=32e^{-|x_1-x_2|}
\eeq
in exact agreement with a formula of Perring and Skyrme, which they found using a direct superposition ansatz
\cite{persky}. Kink-antikink interactions can also be recovered from (\ref{isfi}). Since the antikink
is just a reflected kink, $\phi_{\bar{K}}(x)=\phi_K(-x)$, it
coincides asymptotically with the field induced by a dipole of moment $m=-8$, so
$V_{K\bar{K}}=-32e^{-|x_1-x_2|}$, again in agreement with Perring and Skyrme. So well-separated kinks repel one another, while kinks and antikinks attract one another.

The same basic method works for vortices in the TCGL model, though the details are more complicated. One must
linearize the TCGL model about the  ground state in real $\psi_1$ gauge, but now there are three fields rather than one,
$A$, $\psi_1$ and $\psi_2$, and $\psi_1,\psi_2$ are (generically) directly coupled. The coupling is removed by
expanding $\psi_1,\psi_2$ in a basis of normal modes (eigenvectors of the Hessian of the potential). The linearized
theory is then identified with a pair of uncoupled Klein-Gordon models and a Proca (massive vector boson)
model for $A$. The point source reproducing the asymptotic vortex is a composite with scalar monopole charges for 
the two Klein-Gordon fields and a dipole moment for the vector field $A$.   
{The total interaction energy for
a pair of point vortices is the sum of three terms, two attractive (the scalar monopole interactions) and
one repulsive (the vector dipole interaction), which can be read off from the quadratic terms in the linear 
theory's Lagrangian density, as described in section \ref{asymp}.}


\begin{thebibliography}{99}
\bibitem{remark}
The noninteracting regime, which is frequently called ``Bogomolny limit"
is a property of Ginzburg-Landau model where, at $\kappa = 1/\sqrt{2}$,
the core-core attractive interaction between vortices exactly cancels
the current-current repulsive interaction.
However, in a realistic system extremely close to the
 $\kappa \to 1/\sqrt{2}$ limit one should
go beyond the Ginzburg-Landau description
{to determine leftover inter-vortex interactions, because these are}
then  determined  not by the fundamental length scales of the GL theory but
by small  microscopic effects leading
to non-universal weak vortex interaction potentials.
We do not consider the physics
which arises in the Bogomolny limits in this work.
\end{thebibliography}
\end{document}